\documentclass{llncs}
\usepackage{import} %

\usepackage[utf8]{inputenc}

\PassOptionsToPackage{hyphens}{url}%
\usepackage{hyperref}
\usepackage{graphicx}

\usepackage{amsmath}

\usepackage{amsthm}
\usepackage{amssymb}
\usepackage{amsfonts}
\usepackage{mathtools}

\usepackage{shortsym}

\usepackage{microtype}

\usepackage{IEEEtrantools}
\allowdisplaybreaks

\usepackage[shortlabels,inline]{enumitem}
\setlist[itemize]{leftmargin=1.25em}
\setlist[enumerate]{leftmargin=1.75em}
\usepackage[normalem]{ulem}

\usepackage{import}
\theoremstyle{plain}

\theoremstyle{definition}
\makeatletter
\renewcommand*\env@matrix[1][*\c@MaxMatrixCols c]{%
    \hskip -\arraycolsep
    \let\@ifnextchar\new@ifnextchar
    \array{#1}}
\makeatother

\DeclarePairedDelimiter{\abs}{\lvert}{\rvert}
\DeclarePairedDelimiter{\len}{\lvert}{\rvert}
\DeclarePairedDelimiter{\norm}{\lVert}{\rVert}
\DeclarePairedDelimiter{\floor}{\lfloor}{\rfloor}
\DeclarePairedDelimiter{\ceil}{\lceil}{\rceil}

\makeatletter
\let\oldabs\abs
\def\abs{\@ifstar{\oldabs}{\oldabs*}}
\let\oldlen\len
\def\len{\@ifstar{\oldlen}{\oldlen*}}
\let\oldnorm\norm
\def\norm{\@ifstar{\oldnorm}{\oldnorm*}}
\let\oldfloor\floor
\def\floor{\@ifstar{\oldfloor}{\oldfloor*}}
\let\oldceil\ceil
\def\ceil{\@ifstar{\oldceil}{\oldceil*}}
\makeatother

\usepackage{pifont}

\usepackage{dsfont}

\usepackage{xspace}

\usepackage[dvipsnames]{xcolor}

\definecolor{myA16zGrayLight}{RGB}{235,235,235}     %
\definecolor{myA16zGrayMedium}{RGB}{196,196,196}    %
\definecolor{myA16zGrayDark}{RGB}{44,34,34}         %
\definecolor{myA16zLavender}{RGB}{208,161,255}      %
\definecolor{myA16zMagenta}{RGB}{195,70,206}        %
\definecolor{myA16zMulberry}{RGB}{113,24,88}        %
\definecolor{myA16zLemonChiffon}{RGB}{250,234,157}  %
\definecolor{myA16zAmber}{RGB}{230,154,48}          %
\definecolor{myA16zRust}{RGB}{174,59,10}            %
\definecolor{myA16zLime}{RGB}{197,222,107}          %
\definecolor{myA16zAquamarine}{RGB}{82,216,145}     %
\definecolor{myA16zPine}{RGB}{60,87,44}             %
\definecolor{myA16zPacific}{RGB}{145,224,235}       %
\definecolor{myA16zTeal}{RGB}{36,197,201}           %
\definecolor{myA16zAzure}{RGB}{18,51,90}            %

\definecolor{myTechnionDeepBlue}{HTML}{002147}           %
\definecolor{myTechnionGoldenOchre}{HTML}{D59F0F}        %
\definecolor{myTechnionBlack}{HTML}{000000}              %
\definecolor{myTechnionWhite}{HTML}{FFFFFF}              %

\definecolor{myTechnionRed}{HTML}{E31D1A}             %
\definecolor{myTechnionPink}{HTML}{EA094B}           %
\definecolor{myTechnionPurple1}{HTML}{AE3B72}         %
\definecolor{myTechnionPurple2}{HTML}{4D4084}         %
\definecolor{myTechnionBlue1}{HTML}{216093}            %
\definecolor{myTechnionBlue2}{HTML}{5686DA}           %
\definecolor{myTechnionTeal}{HTML}{32B1CA}            %
\definecolor{myTechnionGreen1}{HTML}{EA094B}           %
\definecolor{myTechnionGreen2}{HTML}{A3D65C}           %
\definecolor{myTechnionGreen3}{HTML}{94D60A}           %
\definecolor{myTechnionYellow}{HTML}{FDD700}          %
\definecolor{myTechnionOrange}{HTML}{FF6B00}         %
\definecolor{myTechnionBrown}{HTML}{97775C}          %
\definecolor{myTechnionBeige}{HTML}{D9D1C3}          %
\definecolor{myTechnionGray1}{HTML}{A2A9AE}            %
\definecolor{myTechnionGray2}{HTML}{5A6771}            %

\definecolor{mySuCardinalRed}{HTML}{8c1515}
\definecolor{mySuCardinalRedLight}{HTML}{B83A4B}
\definecolor{mySuCardinalRedDark}{HTML}{820000}
\definecolor{mySuWhite}{HTML}{ffffff}
\definecolor{mySuCoolGrey}{HTML}{53565A}
\definecolor{mySuBlack}{HTML}{2e2d29}
\definecolor{mySuBlack100}{HTML}{2e2d29}
\definecolor{mySuBlack90}{HTML}{43423E}
\definecolor{mySuBlack80}{HTML}{585754}
\definecolor{mySuBlack70}{HTML}{6D6C69}
\definecolor{mySuBlack60}{HTML}{767674}
\definecolor{mySuBlack50}{HTML}{979694}
\definecolor{mySuBlack40}{HTML}{ABABA9}
\definecolor{mySuBlack30}{HTML}{C0C0BF}
\definecolor{mySuBlack20}{HTML}{D5D5D4}
\definecolor{mySuBlack10}{HTML}{EAEAEA}

\definecolor{mySuPaloAlto}{HTML}{175E54}
\definecolor{mySuPaloAltoLight}{HTML}{2D716F}
\definecolor{mySuPaloAltoDark}{HTML}{014240}
\definecolor{mySuPaloVerde}{HTML}{279989}
\definecolor{mySuPaloVerdeLight}{HTML}{59B3A9}
\definecolor{mySuPaloVerdeDark}{HTML}{017E7C}
\definecolor{mySuOlive}{HTML}{8F993E}
\definecolor{mySuOliveLight}{HTML}{A6B168}
\definecolor{mySuOliveDark}{HTML}{7A863B}
\definecolor{mySuBay}{HTML}{6FA287}
\definecolor{mySuBayLight}{HTML}{8AB8A7}
\definecolor{mySuBayDark}{HTML}{417865}
\definecolor{mySuSky}{HTML}{4298B5}
\definecolor{mySuSkyLight}{HTML}{67AFD2}
\definecolor{mySuSkyDark}{HTML}{016895}
\definecolor{mySuLagunita}{HTML}{007C92}
\definecolor{mySuLagunitaLight}{HTML}{009AB4}
\definecolor{mySuLagunitaDark}{HTML}{006B81}
\definecolor{mySuPoppy}{HTML}{E98300}
\definecolor{mySuPoppyLight}{HTML}{F9A44A}
\definecolor{mySuPoppyDark}{HTML}{D1660F}
\definecolor{mySuSpirited}{HTML}{E04F39}
\definecolor{mySuSpiritedLight}{HTML}{F4795B}
\definecolor{mySuSpiritedDark}{HTML}{C74632}
\definecolor{mySuIlluminating}{HTML}{FEDD5C}
\definecolor{mySuIlluminatingLight}{HTML}{FFE781}
\definecolor{mySuIlluminatingDark}{HTML}{FEC51D}
\definecolor{mySuPlum}{HTML}{620059}
\definecolor{mySuPlumLight}{HTML}{734675}
\definecolor{mySuPlumDark}{HTML}{350D36}
\definecolor{mySuBrick}{HTML}{651C32}
\definecolor{mySuBrickLight}{HTML}{7F2D48}
\definecolor{mySuBrickDark}{HTML}{42081B}
\definecolor{mySuArchway}{HTML}{5D4B3C}
\definecolor{mySuArchwayLight}{HTML}{766253}
\definecolor{mySuArchwayDark}{HTML}{2F2424}
\definecolor{mySuStone}{HTML}{7F7776}
\definecolor{mySuStoneLight}{HTML}{D4D1D1}
\definecolor{mySuStoneDark}{HTML}{544948}
\definecolor{mySuFog}{HTML}{DAD7CB}
\definecolor{mySuFogLight}{HTML}{F4F4F4}
\definecolor{mySuFogDark}{HTML}{B6B1A9}

\definecolor{mySuDigitalRed}{HTML}{B1040E}
\definecolor{mySuDigitalRedLight}{HTML}{E50808}
\definecolor{mySuDigitalRedDark}{HTML}{820000}
\definecolor{mySuDigitalBlue}{HTML}{006CB8}
\definecolor{mySuDigitalBlueLight}{HTML}{6FC3FF}
\definecolor{mySuDigitalBlueDark}{HTML}{00548f}
\definecolor{mySuDigitalGreen}{HTML}{008566}
\definecolor{mySuDigitalGreenLight}{HTML}{1AECBA}
\definecolor{mySuDigitalGreenDark}{HTML}{006F54}

\definecolor{myParula1Blue}{RGB}{0,114,189}
\definecolor{myParula2Orange}{RGB}{217,83,25}
\definecolor{myParula3Yellow}{RGB}{237,177,32}
\definecolor{myParula4Purple}{RGB}{126,47,142}
\definecolor{myParula5Green}{RGB}{119,172,48}
\definecolor{myParula6LightBlue}{RGB}{77,190,238}
\definecolor{myParula7Red}{RGB}{162,20,47}

\usepackage{tikz}
\usetikzlibrary{calc}
\usetikzlibrary{fpu}
\usetikzlibrary{arrows.meta}
\usetikzlibrary{patterns}
\usetikzlibrary{positioning}
\usetikzlibrary{decorations.pathreplacing}
\usetikzlibrary{decorations.pathmorphing}
\usetikzlibrary{calligraphy}
\usetikzlibrary{shapes.misc}
\usetikzlibrary{shapes.symbols}
\usetikzlibrary{shapes.arrows}
\usetikzlibrary{spy}
\usetikzlibrary{shapes.geometric}
\usetikzlibrary{intersections}
\usetikzlibrary{fit}

\usepackage{pgfplots}
\pgfplotsset{compat=1.18}
\usepgfplotslibrary{fillbetween}

\pgfplotsset{
    discard if not/.style 2 args={
            x filter/.code={
                    \edef\tempa{\thisrow{#1}}
                    \edef\tempb{#2}
                    \ifx\tempa\tempb
                    \else
                        
                    \fi
                }
        },
}

\pgfplotsset{
    mysimpleplot/.style = {
            every axis plot/.prefix style={thick},
            width=1.05\linewidth,
            height=0.75\linewidth,
            title style={font=\scriptsize,align=center},
            legend cell align=left,
            legend style={font=\scriptsize},
            legend columns=3,
            legend style={
                    at={(0.5,1)},
                    yshift=0.3em,
                    anchor=south,
                    draw=none,
                    /tikz/every even column/.append style={
                            column sep=0.3em
                        },
                    cells={
                            align=left
                        }
                },
            grid=both,
            minor tick num=9,
            major grid style={solid,very thin,draw=gray!50},
            minor grid style={solid,ultra thin,draw=gray!20},
            label style={font=\scriptsize,align=center},
            tick label style={font=\scriptsize},
        },
}

\pgfplotsset{
    mysimplefig1plot/.style = {
        mysimpleplot,
        xlabel={$\netX$},
        ylabel={$\tALident/n$},
        xmin=0.0, xmax=0.5,
        ymin=0.0, ymax=0.35,
        height=0.75\linewidth,
        width=\linewidth,
        yticklabel style={
                /pgf/number format/fixed,
                /pgf/number format/precision=2
            },
        scaled y ticks=false,
        xtick={0,0.1,0.2,0.25,0.33333,0.5},
        xticklabels={0,1/10,1/5,1/4,1/3,1/2},
        yticklabels={0,1/10,1/6,1/5,1/4,1/3},
        ytick={0,0.1,0.16666,0.2,0.25,0.33333},
    }
}

\tikzset{myparula11/.style={color=myParula1Blue,solid,mark=+,mark options={solid}}}
\tikzset{myparula12/.style={color=myParula1Blue,densely dashed,mark=x,mark options={solid}}}
\tikzset{myparula13/.style={color=myParula1Blue,densely dotted,mark=o,mark options={solid}}}
\tikzset{myparula14/.style={color=myParula1Blue,dashdotted,mark=triangle,mark options={solid}}}
\tikzset{myparula15/.style={color=myParula1Blue,dashdotdotted,mark=square,mark options={solid}}}

\tikzset{myparula21/.style={color=myParula2Orange,solid,mark=+,mark options={solid}}}
\tikzset{myparula22/.style={color=myParula2Orange,densely dashed,mark=x,mark options={solid}}}
\tikzset{myparula23/.style={color=myParula2Orange,densely dotted,mark=o,mark options={solid}}}
\tikzset{myparula24/.style={color=myParula2Orange,dashdotted,mark=triangle,mark options={solid}}}
\tikzset{myparula25/.style={color=myParula2Orange,dashdotdotted,mark=square,mark options={solid}}}

\tikzset{myparula31/.style={color=myParula3Yellow,solid,mark=+,mark options={solid}}}
\tikzset{myparula32/.style={color=myParula3Yellow,densely dashed,mark=x,mark options={solid}}}
\tikzset{myparula33/.style={color=myParula3Yellow,densely dotted,mark=o,mark options={solid}}}
\tikzset{myparula34/.style={color=myParula3Yellow,dashdotted,mark=triangle,mark options={solid}}}
\tikzset{myparula35/.style={color=myParula3Yellow,dashdotdotted,mark=square,mark options={solid}}}

\tikzset{myparula41/.style={color=myParula4Purple,solid,mark=+,mark options={solid}}}
\tikzset{myparula42/.style={color=myParula4Purple,densely dashed,mark=x,mark options={solid}}}
\tikzset{myparula43/.style={color=myParula4Purple,densely dotted,mark=o,mark options={solid}}}
\tikzset{myparula44/.style={color=myParula4Purple,dashdotted,mark=triangle,mark options={solid}}}
\tikzset{myparula45/.style={color=myParula4Purple,dashdotdotted,mark=square,mark options={solid}}}

\tikzset{myparula51/.style={color=myParula5Green,solid,mark=+,mark options={solid}}}
\tikzset{myparula52/.style={color=myParula5Green,densely dashed,mark=x,mark options={solid}}}
\tikzset{myparula53/.style={color=myParula5Green,densely dotted,mark=o,mark options={solid}}}
\tikzset{myparula54/.style={color=myParula5Green,dashdotted,mark=triangle,mark options={solid}}}
\tikzset{myparula55/.style={color=myParula5Green,dashdotdotted,mark=square,mark options={solid}}}

\tikzset{myparula61/.style={color=myParula6LightBlue,solid,mark=+,mark options={solid}}}
\tikzset{myparula62/.style={color=myParula6LightBlue,densely dashed,mark=x,mark options={solid}}}
\tikzset{myparula63/.style={color=myParula6LightBlue,densely dotted,mark=o,mark options={solid}}}
\tikzset{myparula64/.style={color=myParula6LightBlue,dashdotted,mark=triangle,mark options={solid}}}
\tikzset{myparula65/.style={color=myParula6LightBlue,dashdotdotted,mark=square,mark options={solid}}}

\tikzset{myparula71/.style={color=myParula7Red,solid,mark=+,mark options={solid}}}
\tikzset{myparula72/.style={color=myParula7Red,densely dashed,mark=x,mark options={solid}}}
\tikzset{myparula73/.style={color=myParula7Red,densely dotted,mark=o,mark options={solid}}}
\tikzset{myparula74/.style={color=myParula7Red,dashdotted,mark=triangle,mark options={solid}}}
\tikzset{myparula75/.style={color=myParula7Red,dashdotdotted,mark=square,mark options={solid}}}

\usepackage{multirow}
\usepackage{booktabs}   %
\usepackage{makecell}
\usepackage{threeparttable}   %

\usepackage{algorithm}
\usepackage{algorithmicx}
\usepackage[noend]{algpseudocode}

\makeatletter
\AddToHook{env/algorithmic/begin}{\def\@currentcounter{ALG@line}}
\renewcommand{\theHALG@line}{\thealgorithm.\arabic{ALG@line}}
\makeatother

\algnewcommand{\LineComment}[1]{\State {\textcolor{gray}{/\!/ #1}}}

\algrenewcommand{\alglinenumber}[1]{\scriptsize\textcolor{gray}{\texttt{#1}}}
\algrenewcommand{\algorithmicindent}{1em}

\algnewcommand{\algfontsize}[0]{}
\AddToHook{env/algorithmic/begin}{\algfontsize}

\algnewcommand{\algorithmicswitch}{\textbf{switch}}
\algdef{SE}[SWITCH]{Switch}{EndSwitch}[1]{\algorithmicswitch\ #1\ \algorithmicdo}{\algorithmicend\ \algorithmicswitch}%
\algtext*{EndSwitch}%

\algnewcommand{\algorithmiccase}{\textbf{case}}
\algdef{SE}[CASE]{Case}{EndCase}[1]{\algorithmiccase\ #1}{\algorithmicend\ \algorithmiccase}%
\algtext*{EndCase}%

\algnewcommand{\algorithmicon}{\textbf{on}}
\algdef{SE}[ON]{On}{EndOn}[1]{\algorithmicon\ #1\ \algorithmicdo}{\algorithmicend\ \algorithmicon}%
\algtext*{EndOn}%

\algnewcommand{\algorithmicat}{\textbf{at}}
\algdef{SE}[AT]{At}{EndAt}[1]{\algorithmicat\ #1\ \algorithmicdo}{\algorithmicend\ \algorithmicat}%
\algtext*{EndAt}%

\algnewcommand{\algorithmicrealfunction}{\textbf{function}}
\algdef{SE}[REALFUNCTION]{RealFunction}{EndRealFunction}[1]{\algorithmicrealfunction\ #1\ \algorithmicdo}{\algorithmicend\ \algorithmicrealfunction}%
\algtext*{EndRealFunction}%

\algnewcommand{\algorithmicthroughout}{\textbf{do throughout}}
\algdef{SE}[Throughout]{Throughout}{EndThroughout}[1]{\algorithmicthroughout\ #1\ \algorithmicdo}{\algorithmicend\ \algorithmicthroughout}%
\algtext*{EndThroughout}%

\algnewcommand{\algorithmictry}{\textbf{try}}
\algnewcommand{\algorithmiccatch}{\textbf{catch}}
\algdef{SE}[TRY]{Try}{EndTry}{\algorithmictry\ \algorithmicdo}{\algorithmicend\ \algorithmictry}
\algdef{C}[TRY]{TRY}{Catch}[1]{\algorithmiccatch\ #1\ \algorithmicdo}
\algtext*{EndTry}

\algrenewcommand{\algorithmicdo}{}
\algrenewcommand{\algorithmicthen}{}

\algnewcommand{\algorithmicgoto}{\textbf{goto}}%
\algnewcommand{\Goto}[1]{\algorithmicgoto~\ref{#1}}%

\algnewcommand{\algorithmicassert}{\textbf{assert}}%
\algnewcommand{\Assert}[1]{\algorithmicassert~{#1}}%

\algnewcommand{\algorithmicbreak}{\textbf{break}}%
\algnewcommand{\Break}[0]{\algorithmicbreak}%
\algnewcommand{\BreakOutOf}[1]{\algorithmicbreak~out~of~#1}%

\algnewcommand{\algorithmicwaiton}{\textbf{wait on}}%
\algnewcommand{\WaitOn}[1]{\algorithmicwaiton~{#1}}%

\algnewcommand{\InlineRequire}[1]{\textbf{require} {#1}}

\algblock{ManualIndent}{EndManualIndent}
\algnotext{ManualIndent}
\algnotext{EndManualIndent}

\algdef{SE}[GENERICBLOCK]{GenericBlock}{EndGenericBlock}[1]{#1}{}%
\algtext*{EndGenericBlock}%

\usepackage{fontawesome5}
\usepackage{noto-emoji-easy}
\usepackage{tango-icons}

\subimport{./lib/}{defer.tex}
\IfFileExists{version-string.txt}{%
  \newread\versionfile
  \openin\versionfile=version-string.txt
  \read\versionfile to \myVersionString
  \closein\versionfile
}{}

\usepackage[sort&compress,capitalize,nameinlink]{cleveref}

\crefalias{ALG@line}{line}

\crefname{figure}{Fig.}{Figs.}
\Crefname{figure}{Fig.}{Figs.}

\crefname{table}{Tab.}{Tabs.}
\Crefname{table}{Tab.}{Tabs.}

\crefname{section}{Sec.}{Secs.}
\Crefname{section}{Sec.}{Secs.}
\crefname{subsection}{Sec.}{Secs.}
\Crefname{subsection}{Sec.}{Secs.}
\crefname{subsubsection}{Sec.}{Secs.}
\Crefname{subsubsection}{Sec.}{Secs.}
\crefname{subsubsubsection}{Sec.}{Secs.}
\Crefname{subsubsubsection}{Sec.}{Secs.}
\crefname{appendix}{App.}{Apps.}
\Crefname{appendix}{App.}{Apps.}
\crefname{subappendix}{App.}{Apps.}
\Crefname{subappendix}{App.}{Apps.}
\crefname{subsubappendix}{App.}{Apps.}
\Crefname{subsubappendix}{App.}{Apps.}
\crefname{subsubsubappendix}{App.}{Apps.}
\Crefname{subsubsubappendix}{App.}{Apps.}

\crefname{algorithm}{Alg.}{Algs.}
\Crefname{algorithm}{Alg.}{Algs.}
\crefname{line}{ln.}{lns.}
\Crefname{line}{ln.}{lns.}

\crefname{proposition}{Prop.}{Props.}
\Crefname{proposition}{Prop.}{Props.}
\crefname{lemma}{Lem.}{Lems.}
\Crefname{lemma}{Lem.}{Lems.}
\crefname{theorem}{Thm.}{Thms.}
\Crefname{theorem}{Thm.}{Thms.}
\crefname{corollary}{Cor.}{Cors.}
\Crefname{corollary}{Cor.}{Cors.}
\crefname{definition}{Def.}{Defs.}
\Crefname{definition}{Def.}{Defs.}
\crefname{conjecture}{Conj.}{Conjs.}
\Crefname{conjecture}{Conj.}{Conjs.}
\crefname{remark}{Rem.}{Rems.}
\Crefname{remark}{Rem.}{Rems.}

\usepackage[
    sets
]{cryptocode}

\newenvironment{claimproof}[1][\proofname]{%
  \begin{proof}[#1]%
}{%
  \end{proof}%
}

\newcommand{\ip}[1]{\langle #1 \rangle}

\newcommand{\cA}{\ensuremath{\mathcal A}}
\newcommand{\cF}{\ensuremath{\mathcal F}}
\newcommand{\cC}{\ensuremath{\mathcal C}}
\newcommand{\Sync}{\ensuremath{\mathsf{Sync}}}
\newcommand{\DL}{\ensuremath{\mathsf{DL}}}
\newcommand{\bbN}{\ensuremath{\mathbb N}}
\newcommand{\decided}{\ensuremath{\mathsf{decided}}}
\newcommand{\goodcase}{\ensuremath{\mathsf{goodcase}}}
\newcommand{\Tgc}{\ensuremath{T_{\mathrm{gc}}}}

\renewcommand{\BA}{\ensuremath{\mathsf{BA}}}
\renewcommand{\BB}{\ensuremath{\mathsf{BB}}}

\newcommand{\BBfp}{\ensuremath{\Pi_{\BB,2}}}
\newcommand{\BAfp}{\ensuremath{\Pi_{\BA,1}}}
\newcommand{\BBup}{\ensuremath{\Pi_{\BB,\mathsf{UP}}}}
\newcommand{\BBsp}{\ensuremath{\Pi_{\BB,3}}}
\newcommand{\BAsp}{\ensuremath{\Pi_{\BA,2}}}

\newcommand{\BBFP}{\hyperref[prot:BB_fast_path]{\ensuremath{\BBfp}}}
\newcommand{\BAFP}{\hyperref[prot:BA_fast_path]{\ensuremath{\BAfp}}}
\newcommand{\BBUP}{\hyperref[prot:BB_fast_path_unknown_participation]{\ensuremath{\BBup}}}
\newcommand{\BBSP}{\hyperref[prot:BB_slow_path]{\ensuremath{\BBsp}}}
\newcommand{\BASP}{\hyperref[prot:BA_slow_path]{\ensuremath{\BAsp}}}

\newcommand{\echo}{\ensuremath{\mathsf{echo}}}
\newcommand{\vote}{\ensuremath{\mathsf{vote}}}
\newcommand{\forward}{\ensuremath{\mathsf{forward}}}
\newcommand{\echotwo}{\ensuremath{\mathsf{echo2}}}

\usepackage{tcolorbox}
\tcbuselibrary{skins,breakable}

\makeatletter
\newcommand{\DrawLine}{%
  \begin{tikzpicture}
  \path[use as bounding box] (0,0) -- (\linewidth,0);
  \draw[color=black!75,dashed,dash phase=2pt]
        (0-\kvtcb@leftlower-\kvtcb@boxsep,0)--
        (\linewidth+\kvtcb@rightlower+\kvtcb@boxsep,0);
  \end{tikzpicture}%
  }
\makeatother

\newtcolorbox{mybox}[2][]{%
  enhanced,
  title        = {#2},
  attach boxed title to top left={xshift=+3mm,yshift*=-3mm},
  breakable    = true,
  colback      = black!4,
  colframe     = black!75,
  fonttitle    = \bfseries,
  colbacktitle = black!10!white,
  coltitle     = black,
  #1
}

\newtcolorbox[auto counter]{functionality}[2][]{%
  enhanced,
  title        = {Functionality~\thetcbcounter: #2},
  attach boxed title to top left={xshift=+3mm,yshift*=-3mm},
  breakable    = true,
  colback      = yellow!4,
  colframe     = black!75,
  fonttitle    = \bfseries,
  fontupper    = \small,
  fontlower    = \small,
  colbacktitle = yellow!10!white,
  coltitle     = black,
  #1
}

\newtcolorbox[use counter=pro]{protocol}[2][]{%
  enhanced,
  title        = {Protocol~\thetcbcounter: #2},
  attach boxed title to top left={xshift=+3mm,yshift*=-3mm},
  breakable    = true,
  colback      = black!4,
  colframe     = black!75,
  fonttitle    = \bfseries, 
  fontupper    = \small,
  fontlower    = \small,
  colbacktitle = black!10!white,
  coltitle     = black,
  #1
}

\newtcbox{\xmybox}[1][red]{on line,
arc=7pt,colback=#1!10!white,colframe=#1!50!black,
before upper={\rule[-3pt]{0pt}{10pt}},boxrule=1pt,
boxsep=0pt,left=6pt,right=6pt,top=2pt,bottom=2pt}

\title{Optimal Good-Case Latency\\for Sleepy Consensus}

\author{
Yuval Efron\inst{1} \and
Joachim Neu\inst{2} \and
Ling Ren\inst{3} \and
Ertem Nusret Tas\inst{2}
}

\institute{
Columbia University, \email{ye2210@columbia.edu}
\and
a16z Crypto Research, \email{\{jneu,ntas\}@a16z.com}
\and
University of Illinois at Urbana--Champaign, \email{renling@illinois.edu}
}

\date{\today}
\begin{document}
\maketitle
\begin{abstract}
In the context of Byzantine consensus problems such as Byzantine broadcast (BB) and Byzantine agreement (BA), the \emph{good-case} setting aims to study the minimal possible latency of a BB or BA protocol under certain favorable conditions, namely the designated leader being correct (for BB), or all parties having the same input value (for BA).
We provide a full characterization of the feasibility and impossibility of good-case latency, for both BA and BB, in the synchronous \emph{sleepy} model.
Surprisingly to us, we find \emph{irrational} resilience thresholds emerging:
$2$-round good-case BB is possible 
if and only if 
at all times,
at least $\frac{1}{\varphi} \approx 0.618$ fraction of the active parties
are correct, where $\varphi = \frac{1+\sqrt{5}}{2} \approx 1.618$ is the golden ratio;
$1$-round good-case BA is possible
if and only if
at least $\frac{1}{\sqrt{2}} \approx 0.707$ fraction of the active parties are correct.
\end{abstract}

\section{Introduction}
\label{sec:intro}

The Byzantine consensus problems~\cite{lamport2019byzantine}, such as Byzantine broadcast and Byzantine agreement, are among the most fundamental and well-studied problems in cryptography and distributed computing. The recent surge of cryptocurrencies and blockchain systems has rekindled interest in these problems, and also drawn the community's attention to unconventional system models that were overlooked in the decades of prior research. 

One of the most prominent and striking features of the celebrated Nakamoto consensus protocol~\cite{Bitcoin,bitcoinbackbone} 
is that it allows correct parties to participate intermittently in the protocol. 
At any time, only a (possibly tiny) fraction of correct parties are actively following the protocol. The rest of the correct parties are in an \emph{inactive} (or crashed) state. Furthermore, correct parties can transition between active and inactive states at any time without prior notice, and they do not know for sure how many other parties are actively participating in the protocol at any time.
A variant of this model (without reliance on proof-of-work) is formalized by Pass and Shi as the \emph{sleepy} model~\cite{sleepy}, also interchangeably called the \emph{dynamic participation} model.\footnote{Importantly, the \emph{unexpected temporary crash faults} (what ``dynamic participation'' refers to) in the sleepy model is not to be confused with \emph{scheduled} reconfiguration of the membership set of parties operating the protocol (also called ``stake shift'' in the context of proof-of-stake blockchains). No reconfiguration of the set of operating parties takes place throughout this paper.}
In contrast, we refer to the traditional model where all correct parties are always active as the \emph{static participation} model.
Note that a variant of the CAP theorem~\cite{captheorem,blockchaincap,ebbandflow} shows that a synchronous network is necessary for sleepy consensus, so we assume \emph{synchrony} throughout this paper.
The sleepy model is employed by leading blockchains such as Ethereum and Cardano~\cite{ebbandflow,Goldfish,ouroboros,DBLP:conf/ccs/BadertscherGKRZ18}.

A line of recent works has designed sleepy Byzantine consensus protocols that match static participation protocols in many aspects. 
Sleepy Byzantine consensus protocols can tolerate minority Byzantine faults~\cite{ouroboros,sleepy,GLR21,MR22,Losa23,GL23,MMR23,Goldfish,DACS,fullyfluctuating,DZ23b}, 
which is the same corruption threshold in the static model.
Protocols in the sleepy model have also achieved expected constant rounds~\cite{MR22,MMR22,Losa23,GL23,MMR23,DACS,fullyfluctuating},
again matching the static participation model.

In this paper, we focus on the \emph{good-case} latency $\Tgc$, measured in rounds of communication.
For the Byzantine broadcast problem, the ``good case'' refers to the case where the designated sender is correct, while for Byzantine agreement, it refers to the case where all parties have the same input value.
In addition to a certain latency requirement in the good case, the protocols are required to satisfy the usual consensus desiderata (termination, agreement, validity; see \cref{def:BB,def:BA}) even when the good-case conditions are not met.
The good-case latency is a metric well-motivated by practical considerations.
Most deployed consensus protocols have parties rotate to serve as the \emph{leader}, who proposes a value (a block) for other parties to agree on.
Because there are mechanisms to reward or penalize the leader based on its actions, the vast majority of leaders behave correctly. 
We thus would like the consensus protocol to be as fast as possible under a correct leader.

\begin{table}[tbp]
    \centering
    \caption{Achievable and impossible resiliences for Byzantine broadcast (BB) and Byzantine agreement (BA) with varying good-case latency $\Tgc$ in different participation models.
    Notably, we provide a tight characterization under dynamic participation, where tight irrational resilience thresholds emerge: $1-1/\varphi \approx 0.382$ (where $\varphi = \frac{1+\sqrt{5}}{2} \approx 1.618$ is the golden ratio) and $1-1/\sqrt{2} \approx 0.293$.}
    \label{tab:results2}
    \vspace{-0.5em}
    \begin{threeparttable}
        \begin{tabular}{@{\hspace{0.5em}}l@{\hspace{1.5em}}c@{\hspace{1.5em}}l@{\hspace{1.5em}}l@{\hspace{1.5em}}l@{\hspace{0.5em}}}
            \toprule
            \textbf{Problem} & \textbf{$\Tgc$} & \textbf{Part.\ Model} & \textbf{Achiev.\ Resil.} & \textbf{Imposs.\ Resil.} \\
            \midrule
            BB & $2$ & Static P.  & $1/2$ \cite{AbrahamN0X21} & Open\tnote{$\ast$} \\
               &     & Unknown P. & $1/2$ (\cref{thrm:BB_fast_path_unknown_participation}) & Open\tnote{$\ast$} \\
               &     & Dynamic P.  & $1-1/\varphi$ (\cref{thrm:BB_fast_path}) & $1-1/\varphi$ (\cref{thrm:good-case-BB-3-rounds}) \\
            \midrule
            BB & $\geq 3$ & Static P.  & $1/2$ \cite{AbrahamN0X21}\tnote{$\dagger$} & Open\tnote{$\ast$} \\
               &     & Unknown P. & $1/2$ (\cref{thrm:BB_fast_path_unknown_participation})\tnote{$\dagger$} & Open\tnote{$\ast$} \\
               &     & Dynamic P. & $1/2$ (\cref{thrm:BB_slow_path}) & $1/2$ \cite{adversarymajority} \\
            \midrule
            BA & $1$ & Static P.  & $1/3$ (Folk.: \cref{lemma:one_round_sync_BA}) & $1/3$ (Folk.: \cref{lemma:one_round_sync_BA_lower_bound}) \\
               &     & Unknown P. & $1-1/\sqrt{2}$ (\cref{thrm:BA_fast_path,})\tnote{$\ddagger$} & $1-1/\sqrt{2}$ (\cref{thrm:good-case-BA-2-rounds}) \\
               &     & Dynamic P. & $1-1/\sqrt{2}$ (\cref{thrm:BA_fast_path}) & $1-1/\sqrt{2}$ (\cref{thrm:good-case-BA-2-rounds})\tnote{$\ddagger$} \\
            \midrule
            BA & $\geq 2$ & Static P.  & $1/2$ (\cref{thrm:BA_slow_path})\tnote{$\ddagger$} & $1/2$ \cite{bahalfresilience} \\
               &     & Unknown P. & $1/2$ (\cref{thrm:BA_slow_path})\tnote{$\ddagger$} & $1/2$\tnote{$\ddagger$} \\
               &     & Dynamic P. & $1/2$ (\cref{thrm:BA_slow_path}) & $1/2$\tnote{$\ddagger$} \\
            \bottomrule
        \end{tabular}
        \begin{tablenotes}
            \item 
                ``Folk.'' indicates a result is folklore.
            \item[$\ast$] The landscape of Byzantine broadcast in the ``adversarial majority'' regime remains incomplete, even under static participation.
            The protocol provided in~\cite{shidishonestmajority} achieves $\Tgc = O(\frac{n}{n-f})$. 
            This matches the impossibility results~\cite{AbrahamN0X21} asymptotically, but concrete gaps remain.
            This lack of clarity seeps into the unknown participation model,
            since crashed honest parties can be treated as Byzantine for the purpose of protocols tolerating an adversarial majority of parties~\cite{adversarymajority}.
            For dynamic participation, we fully characterize the good-case latency and resilience achievable for Byzantine broadcast and Byzantine agreement.
            \item[$\dagger$] Implied by a protocol with better latency.
            \item[$\ddagger$] Implied by a protocol or impossibility result for a more demanding model.
        \end{tablenotes}
    \end{threeparttable}
\end{table}

\begin{table}[tbp]
    \centering
    \caption{Earlier protocols and their good-case latency $\Tgc$ and resilience, for Byzantine broadcast (BB) and Byzantine agreement (BA), in different participation models.}
    \label{tab:relatedworks}
    \vspace{-0.5em}
    \begin{threeparttable}
        \begin{tabular}{@{\hspace{0.5em}}l@{\hspace{1.5em}}l@{\hspace{1.5em}}l@{\hspace{1.5em}}c@{\hspace{1.5em}}c@{\hspace{0.5em}}}
            \toprule
            \textbf{Problem} & \textbf{Part.\ Model} & \textbf{Protocol} & \textbf{Lat.\ $\Tgc$} & \textbf{Resil.} \\
            \midrule
            Broadcast (BB) & Static P.  & ANRX'21~\cite{AbrahamN0X21} & $2$ & $1/2$ \\
                           & Unknown P. & KW'21~\cite{KW21} & $3$ & $1/3$ \\
                           & Dynamic P. & MMR'23~\cite{MMR23} & $4$ & $1/2$ \\
            \midrule
            Agreement (BA) & Static P.  & Folklore (\cref{lemma:one_round_sync_BA}) & 1 & $1/3$ \\
                           & Static P.  & MR'21~\cite{MomoseR21} & $4$ & $1/2$ \\
                           & Unknown P. & KW'21~\cite{KW21} & $3$ & $1/3$ \\
                           & Dynamic P. & LG'23~\cite{Losa23,GL23} & $4$ & $1/2$ \\
                           & Dynamic P.\tnote{$\dagger$} & DSTZ'23~\cite{DAmato23} & $3$ & $1/2$  \\
            \bottomrule
        \end{tabular}
        \begin{tablenotes}
            \item[$\dagger$] Mild stable participation assumption.
        \end{tablenotes}
    \end{threeparttable}
\end{table}

Note that no Byzantine broadcast protocol can hope to achieve latency $\Tgc \leq 1$, even in the good case, for the straightforward reason that it leaves correct parties no time to reconcile (or even detect) equivocating values they may have received from a Byzantine leader in the first communication round.
Similarly, no Byzantine agreement protocol can hope to achieve good-case latency $\Tgc = 0$, for the analogous reason that it leaves correct parties no time to reconcile different input values they may have.
On the other hand, an earlier work~\cite{AbrahamN0X21} has shown that for Byzantine broadcast in the traditional synchronous model with static participation, two rounds are sufficient and necessary in the good case, if the \emph{resilience} $\rho$ (i.e., fraction of Byzantine parties tolerable) is below $1/2$ (see \cref{tab:results2}).
For Byzantine agreement in the same model, it is folklore that one round latency can be achieved in the good case, if and only if at most $1/3$ of parties are Byzantine (see \cref{tab:results2,lemma:one_round_sync_BA,lemma:one_round_sync_BA_lower_bound}).

\paragraph{Results.}
This paper extends the systematic study of good-case latency of Byzantine consensus to the dynamic participation (sleepy) model.
We give a complete characterization of the good-case latency of Byzantine broadcast and Byzantine agreement, and the corresponding resilience (see \cref{tab:results2}).
Our results for the dynamic participation model imply a complete characterization of good-case latency and resilience for Byzantine agreement and Byzantine broadcast (in the ``honest majority'' regime) in the \emph{unknown participation} model~\cite{KW21}.
The unknown participation model is weaker than the traditional static participation model, but stronger than the dynamic participation model, in the sense that the number of active correct parties is unknown to the protocol, but does not change once the protocol starts.

Perhaps the most surprising part of our findings is that the critical resilience thresholds that dictate good-case latency under dynamic and unknown participation are \emph{irrational} numbers.
For instance, Byzantine broadcast with good-case latency $\Tgc=2$ and resilience $\rho$ can be achieved under dynamic participation if and only if $\rho \leq 1-1/\varphi \approx 0.382$, where $\varphi = \frac{1+\sqrt{5}}{2} \approx 1.618$ is the golden ratio.
Byzantine agreement with good-case latency $\Tgc=1$ and resilience $\rho$ can be achieved under dynamic participation and under unknown participation if and only if $\rho \leq 1-1/\sqrt{2} \approx 0.293$.
To the best of our knowledge, this is the first time irrational corruption thresholds arise in distributed and threshold cryptography problems, such as Byzantine consensus, verifiable secret sharing, and multi-party computation. 

Our results also establish separations among static, unknown, and dynamic participation.
When the corruption threshold is between $1-1/\varphi$ and $1/2$, Byzantine broadcast under unknown participation can decide in two rounds in the good case, but requires three rounds under dynamic participation.
When the target corruption threshold is between $1-1/\sqrt{2}$ and $1/3$, Byzantine agreement under static participation can decide in one round in the good case, but requires two rounds under unknown participation.

\paragraph{Related Work.}
\Cref{tab:relatedworks} lists earlier consensus protocols with low good-case latency.
For Byzantine broadcast,~\cite{AbrahamN0X21} achieves the optimal $\Tgc=2$ under static participation.
For dynamic participation,~\cite{MMR23} provides $\Tgc=4$ under a majority of correct active parties,
which also works under unknown participation, whereas~\cite{KW21} gives a protocol with improved latency $\Tgc=3$, but only with resilience $1/3$.
We provide broadcast protocols for dynamic participation with minimal good-case latency, $\Tgc=2$ and $\Tgc=3$, and optimal resilience, $1-1/\varphi$ and $1/2$, respectively (\cref{tab:results2}).

For Byzantine agreement, a folklore result (see \cref{lemma:one_round_sync_BA,lemma:one_round_sync_BA_lower_bound}) achieves the minimal $\Tgc=1$ with $1/3$ resilience despite synchrony and static participation.
Under dynamic participation,~\cite{Losa23,GL23} achieves $\Tgc=4$ under $1/2$ resilience,
while~\cite{DAmato23} improves it to $\Tgc=3$ at the expense of a mild stable-participation assumption.
The work of~\cite{DAmato23} also implies protocols with $\Tgc=3$ and $1/2$ resilience under unknown and static participation models.
We provide agreement protocols for dynamic participation (and for unknown participation) with minimal good-case latency, $\Tgc=1$ and $\Tgc=2$, and optimal resilience, $1-1/\sqrt{2}$ and $1/2$, respectively (\cref{tab:results2}).

\paragraph{Outline.}
We briefly review the model and definitions in \cref{sec:model_and_definitions},
before providing a technical overview of our results in \cref{sec:technical_overview}.
Then,
\Cref{sec:lower_bounds} proves our impossibility results.
\Cref{sec:fast_path_BB_BA} presents our BB and BA protocols with $\Tgc=2$ and $\Tgc=1$, respectively, and proves their correctness, before \cref{sec:slow_path_BB_BA} does the same for our BB and BA protocols with $\Tgc=3$ and $\Tgc=2$, respectively.

\section{Model \& Definitions}
\label{sec:model_and_definitions}

\subsection{Model}
\label{ssec:model}

\paragraph{Sleepy Model of Consensus.} 
We work in the \emph{sleepy} model~\cite{sleepy} of Pass and Shi, also called the \emph{dynamic participation} model.
We consider a setting with $N$ parties in total, identified by a public key infrastructure (PKI). We assume a synchronous pairwise communication 
network, with a known delay upper-bound of $\Delta$ on message delivery. For simplicity, we present both our upper and lower bounds for the case $\Delta=1$, though both generalize to general $\Delta$ in a straightforward manner. We consider some known universe $I$ of possible input values.

\paragraph{The Adversary.}
We consider a \emph{static corruption} adversary $\cA$, i.e., at round $t=0$, the adversary chooses a set $F\subseteq [N]$ of parties to be \emph{corrupt}. The remaining parties are \emph{correct}. Corrupt parties are entirely controlled by the adversary, and may deviate arbitrarily from the protocol.
Furthermore, following~\cite{sleepy}, at all rounds $t$, the adversary can adaptively\footnote{The adaptive adversary is not strongly rushing, i.e., it \emph{cannot} observe the contents of a message sent by a party, intercept it, and then put the party to sleep.} select a subset of the correct parties to be \emph{asleep} (i.e., \emph{inactive}), whereas the remaining correct parties are \emph{awake} (i.e., \emph{active}).
Asleep parties are temporarily crashed: they do not execute the protocol or send messages.
Messages sent to an asleep party are buffered and delivered to the party (in an adversarially selected order) the next time it wakes up. 
Upon waking up, parties know the current round number, i.e., we assume synchronized clocks. 
Note that corrupt parties are awake from round $t=0$, and never go to sleep throughout the execution.
The number of awake parties at round $t$ is denoted by $n_t$. We drop the subscript when the round is clear from the context. 
For an adversary $\cA$, we denote by $F_\cA$ the set of parties corrupted by $\cA$, and we usually omit the subscript when the adversary is clear from context. For a value $\rho\in [0,1]$, we say that an adversary $\cA$ is $\rho$-bounded if for all $t$, $|F|<\rho n_t$.
Finally, whenever relevant, the adversary chooses the input values (from $I$) of all correct parties.

\paragraph{Unknown Participation Model.} 
Some of our results pertain to the unknown participation model, introduced in~\cite{KW21}. Unknown participation is a special case of the dynamic participation model, which can formally be described by
a restricted family of adversaries, as follows. Consider the exact same setup as above, with the following restriction on the adversary model: At time $t=0$, the adversary picks a set $S\subseteq [N]$ of $n$ awake parties and a set $F\subseteq S$ of corrupt parties. From this point, the adversary cannot wake up or put to sleep any party. In other words, the parties in $S\backslash F$ are correct and awake throughout the entire execution, the parties in $F$ are corrupt and awake throughout the entire execution, and the remaining parties are correct and asleep throughout the entire execution. 
Similarly to the dynamic participation adversary, we say that an adversary $\cA$ is $\rho$-bounded if $|F|<\rho n$.

\paragraph{Cryptographic Primitives.} 
We use a digital signature scheme on top of our PKI assumption. We assume $\cA$ is computationally bounded, i.e., $\cA$ cannot forge signatures on any message on behalf of parties outside of $F_\cA$.

\subsection{Consensus Primitives}
\label{ssec:consensus_primitives}

We consider two standard consensus problems in this paper, Byzantine broadcast (BB) and Byzantine agreement (BA).
\begin{definition}[Byzantine Broadcast ($\BB$)]
    \label{def:BB}
    In the $\BB$ problem, there is a designated party $\ell$, referred to as the \emph{leader}, with an \emph{input} $u\in I$. As output, each party $p$ \emph{decides} a value $o_p\in I$. Let $\Pi$ be a protocol to be executed by the parties. The following properties constitute the $\BB$ problem. 
    \begin{enumerate}
        \item \textbf{Termination.} There exists a round $T$ such that every correct party decides and halts in the first round $T'\geq T$ in which it is awake.
        \item \textbf{Agreement.} There exists a value $u'\in I$ such that all correct parties that decide, decide $u'$.
        \item \textbf{Validity.} If $\ell$ is correct and awake in round $t=0$, then all correct parties that decide, decide $u$, i.e., $\ell$'s input.
    \end{enumerate}
    For a value $\rho$, we say that a protocol $\Pi$ for $\BB$ is $\rho$-secure under dynamic/unknown participation if it satisfies termination, agreement, and validity against any $\rho$-bounded adversary.
    The maximum $\rho$ for which $\Pi$ is $\rho$-secure is $\Pi$'s \emph{resilience}.
\end{definition}

\begin{definition}[Byzantine Agreement ($\BA$)]
    \label{def:BA}
    In the $\BA$ problem, each party $p$ has an \emph{input} $u_p\in I$. As output, each party $p$ \emph{decides} a value $o_p\in I$. Let $\Pi$ be a protocol to be executed by the parties. The following properties constitute the $\BA$ problem. 
    \begin{enumerate}
        \item \textbf{Termination.} There exists a round $T$ such that every correct party decides and halts in the first round $T'\geq T$ in which it is awake.
        \item \textbf{Agreement.} There exists a value $u'\in I$ such that all correct parties that decide, decide $u'$.
        \item \textbf{Validity.} If all correct parties that are awake in round $t=0$ have $u$ as input, then all correct parties that decide, decide $u$. 
    \end{enumerate}
    For a value $\rho$, we say that a protocol $\Pi$ for $\BA$ is $\rho$-secure under dynamic/unknown participation if it satisfies termination, agreement, and validity against any $\rho$-bounded adversary.
    The maximum $\rho$ for which $\Pi$ is $\rho$-secure is $\Pi$'s \emph{resilience}.
\end{definition}

\subsection{Good-Case Latency}
\label{ssec:goodcase_latency}

\paragraph{Decision Latency.} 
For a $\rho$-bounded adversary $\cA$ and a $\rho$-secure protocol $\Pi$ for $\BB$ or $\BA$, we denote by $\DL_\cA(\Pi)$ the random variable (over the randomness used by correct parties and the adversary) indicating the smallest round $R$ for which every correct party \emph{decides} (but not necessarily terminates) in the first round $R'\geq R$ in which it is awake, in the presence of $\cA$. We say that $\Pi$ has decision latency $R$ with respect to a $\rho$-bounded adversary $\cA$ if $R=\arg\min\limits_{R'\in \bbN}\Pr[\DL_\cA(\Pi)\leq R']=1$. For a family $\cF$ of $\rho$-bounded adversaries we define the decision latency of $\Pi$ with respect to $\cF$ to be $\DL_\cF(\Pi)=\sup\limits_{\cA\in \cF} \DL_\cA(\Pi)$.

\paragraph{Good-Case Latency.} 
The efficiency metric we focus on in this work is that of \emph{good-case} decision latency $\Tgc$. Intuitively, this measure considers the decision latency of a protocol $\Pi$ in executions in which certain favorable conditions hold, namely $\ell$ being correct for $\BB$, or all correct parties having the same input for $\BA$. More formally, we define good-case latency as follows.
\begin{enumerate}
    \item $\BB$: For a value $\rho$, and a $\rho$-secure protocol $\Pi$ for $\BB$ under dynamic/unknown participation, consider the family $\cF_{\goodcase}$ of $\rho$-bounded adversaries $\cA$ for which the leader $\ell$ is correct, i.e., $\ell\not\in F$, and the leader is awake at round $t=0$. We define the good-case latency of $\Pi$ to be $\Tgc \triangleq \DL_{\cF_{\goodcase}}(\Pi)$.
    \item $\BA$: For a value $\rho$, and a $\rho$-secure protocol $\Pi$ for $\BA$ under dynamic/unknown participation, consider the family $\cF_{\goodcase}$ of $\rho$-bounded adversaries $\cA$ for which $\cA$ gives all correct parties awake at round $t=0$ the same input value $u$, for some $u\in I$. We define the good-case latency of $\Pi$ to be $\Tgc \triangleq \DL_{\cF_{\goodcase}}(\Pi)$.
\end{enumerate}

\section{Technical Overview}
\label{sec:technical_overview}

\paragraph{The Traditional Case of Synchrony with Static Participation.} 
To develop some intuition about the feasibility of good-case latency, let us begin by considering the traditional synchronous model with static participation, which in our formulation translates to adversaries for which all correct parties are awake in all rounds. Pondering on the $\BB$ problem for a moment, it definitely seems like in $2$ rounds, we cannot do much beyond:
\begin{enumerate}
    \item Round 0: Have the leader $\ell$ multicast its value.
    \item Round 1: Have all parties report what they heard from the leader.
\end{enumerate}

If the total number of parties is $N$, keeping this protocol blueprint in mind, for some value $\rho\in [0,\frac{1}{2}]$, we can expect correct parties to collect strictly more than $(1-\rho) N$ amount of ``evidence'' for a value $u$ against $\rho$-bounded adversaries, in executions in which $\ell$ is correct with input $u$.
An additional observation is that in executions with a correct leader $\ell$, correct parties can also expect to collect $0$ evidence for any other value. As such we can have a correct party decide after two rounds if it received sufficient evidence for a value, and \emph{no} evidence for any other value. One can observe that this implies that if a party $p$ decided $u$, then any other correct party $q$ observed $<\rho N$ evidence for any other value. To allow other parties to break the symmetry, all we need to stipulate is that $(1-\rho)N\geq \rho N$. Note that this bound holds for all $\rho\leq \frac{1}{2}$, and indeed $\rho=\frac{1}{2}$ turns out to be the correct corruption bound for the feasibility of 2-round good-case $\BB$ in the traditional synchronous static participation setting.

\paragraph{Generic Intuition.} 
Given that this approach works for the static participation case, let us test whether the following line of thinking can yield the correct corruption bound for 2-round good-case BB in other settings: 
\emph{For a bound $\rho$, derive a lower bound on the amount of evidence $c_\rho$ available for the leader's input in executions with a correct leader. Given this amount, derive an upper bound on the amount of evidence $a_\rho$ a correct party can receive for a different value. Stipulate $c_\rho\geq a_\rho$.}

\paragraph{Dynamic Participation.}
Let us apply the above intuition to dynamic participation. For a bound $\rho$, consider a good-case execution (Execution 1), i.e., the leader is correct with some value $u$ and awake at round $t=0$, for a $\rho$-bounded adversary $\cA$. In a similar fashion to the previous paragraph, consider a correct party $p$ awake in round $t=2$, and suppose it hears from $n_p$ parties. It can expect to hear evidence for $u$ from at least $c_\rho=(1-\rho)n_p$ of them, and no evidence for any other value. From the perspective of $p$, the $\rho n_p$ parties reporting no evidence for any value are corrupt and $p$ should ignore them and decide anyway. One might be tempted now to repeat the argument from the previous paragraph, and claim that any other correct party can receive at most $\rho n_p$ evidence for any other value, and so we again have $(1-\rho)n_p\geq \rho n_p$ yielding $\rho=\frac{1}{2}$. 

The key observation is that \emph{$p$ does not know that the parties reporting no evidence are actually corrupt}, as it cannot be certain about the total number of participating parties. To make this concrete, consider now an alternative case (Execution 2), in which the leader is corrupt, and although $p$ hears from $n_p$ parties, they are all correct, and there are in fact $\frac{\rho}{1-\rho}n_p$ corrupt parties, and not just $\rho n_p$. The corrupt leader does not send anything to $(1-\rho)n_p$ of the correct parties, and the rest of the corrupt parties do not send anything to $p$. From $p$'s perspective at round $t=2$, execution 2 is identical to execution 1, and so $p$ decides in round $t=2$ in both of them. In execution 2, however, a different correct party may receive $a_\rho=\frac{\rho}{1-\rho}n_p$ evidence for a value $u'\neq u$. Thus to allow other parties to break the symmetry, we must stipulate $(1-\rho)n_p\geq \frac{\rho}{1-\rho}n_p$, which yields $\rho\leq (1-\rho)^2$, which in turn gives $\rho\leq 1-\frac{1}{\varphi}$, where $\varphi$ is the golden ratio! 

This intuition turns out to yield the correct solution, and we provide tight lower and upper bounds in \cref{thrm:good-case-BB-3-rounds}, and \cref{thrm:BB_fast_path}, respectively. We complement these results by establishing in \cref{thrm:BB_slow_path} that 3-round good-case $\BB$ is feasible under dynamic participation for $\rho=\frac{1}{2}$.

\paragraph{Unknown Participation.} 
An astute reader may rush to a premature conclusion upon reading the previous part, as it seems that the same intuition from above should apply to and yield the same bound under unknown participation. At least at first glance, it seems that the only thing we used is $p$'s inability to distinguish between the case of $n_p$ participating parties and $\frac{n_p}{1-\rho}$ participating parties. Perhaps, however, we were too hasty in making that assertion. 

This intuition turns out to be deceiving under unknown participation. Specifically, $p$ can more scrupulously rely on the fact that \emph{the same set of correct parties is awake in all rounds}. In protocol terms, parties can multicast a \emph{heartbeat} message both in round $t=0$ and in round $t=1$, along with the leader's proposed value; they also relay all received heartbeat messages. Now, a protocol can stipulate that having $q$'s heartbeat relayed by a majority of the parties is a prerequisite for $p$ to count $q$'s contribution to the leader's purported value. It turns out that this technique allows the unknown participation model to boast 2-round good-case $\BB$ for $\rho=\frac{1}{2}$, thus establishing a separation between dynamic and unknown participation. The formal details can be found in \cref{thrm:BB_fast_path_unknown_participation}.

\paragraph{$\BA$.} 
While for $\BB$, the good-case decision latency is either 2 rounds to 3 rounds, the good-case decision latency of $\BA$ can be as small as a single round. It turns out that 2-round good-case $\BA$ is feasible for $\rho=\frac{1}{2}$ even under dynamic participation!
To the best of our knowledge, this result was not known even in the static participation setting, even though it is rather straightforward in that model. For dynamic participation, however, such a result is far from trivial, and involves a novel \emph{report participation} technique. See \cref{thrm:BA_slow_path} for the full details.

This leaves us with the question: what values of $\rho$ make 1-round good-case $\BA$ feasible? The same generic intuition we applied for $\BB$ under dynamic participation can be repeated, with the caveat that we now have to account for equivocations by corrupt parties when calculating $a_\rho$. The math yields that we must stipulate $\rho\leq (1-\rho)^2-\rho(1-\rho)$, which is tight for $\rho=1-\frac{1}{\sqrt{2}}$. The lower and upper bounds can be found in \cref{thrm:good-case-BA-2-rounds}, and \cref{thrm:BA_fast_path}, respectively. In fact, our $\BA$ lower bound (\cref{thrm:good-case-BA-2-rounds}) applies even for unknown participation (the easier setting), while our $\BA$ upper bound (\cref{thrm:BA_fast_path}) applies to dynamic participation (the harder setting). Furthermore, this establishes a separation between the static participation model, for which 1-round good-case $\BA$ is feasible for $\rho\leq \frac{1}{3}$, and the unknown participation model.

\section{Lower Bounds}
\label{sec:lower_bounds}

In this section, we prove the following theorems.
\begin{theorem}\label{thrm:good-case-BB-3-rounds}
    Let $\varphi = \frac{1+\sqrt{5}}{2} \approx 1.618$ be the golden ratio.
    For all $\rho > 1 - \frac{1}{\varphi} \approx 0.382$,
    there is no $\rho$-secure protocol solving $\BB$ under dynamic participation with good-case latency $\leq 2$. 
\end{theorem}

\begin{theorem}\label{thrm:good-case-BA-2-rounds}
    For all $\rho > 1-\frac{1}{\sqrt{2}} \approx 0.293$,
    there is no $\rho$-secure protocol solving $\BA$ under unknown participation with good-case latency $\leq 1$. 
\end{theorem}

\subsection{BB Lower Bound}
\label{ssec:BB_lower_bound}
In this section, we prove \cref{thrm:good-case-BB-3-rounds}.

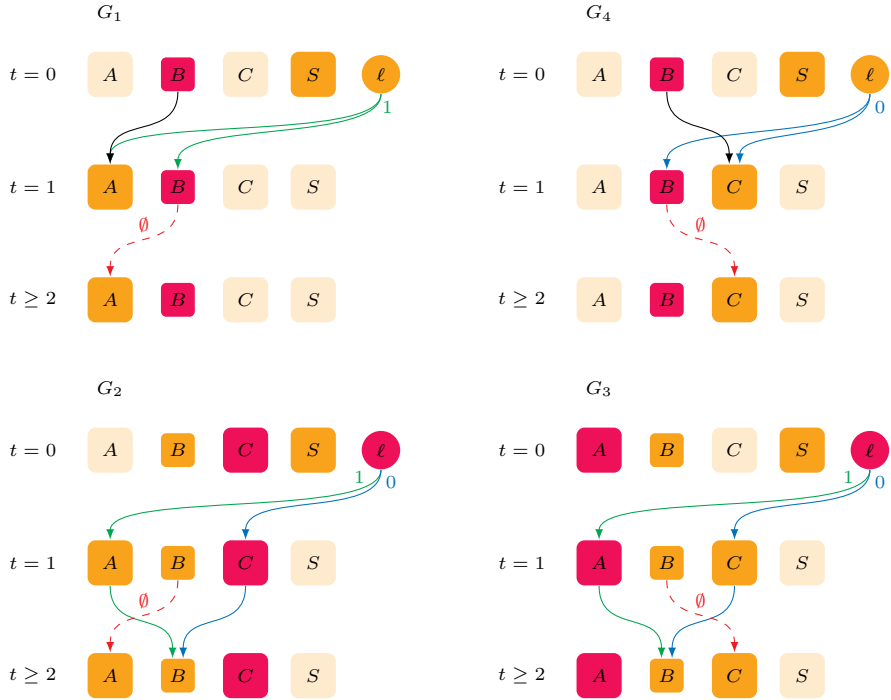
\begin{figure}[tbp]
    \centering
    \begin{tikzpicture}[
        x=0.9cm,
        y=1.5cm,
        leader/.style={shape=circle},
        node/.style={},
        big/.style={minimum width=0.6cm, minimum height=0.6cm, rounded corners=3pt},
        sml/.style={minimum width=0.45cm, minimum height=0.45cm, rounded corners=2pt},
        honact/.style={fill=YellowOrange},
        honslp/.style={fill=YellowOrange!20},
        adv/.style={fill=OrangeRed},
        sim0/.style={-latex,color=RoyalBlue},
        sim1/.style={-latex,color=Green},
        simedge/.style={out=-90,in=90,out distance=2em,in distance=2em},
        omit/.style={-latex,dashed,color=Red},
        correct/.style={-latex,color=black},
        execlabel/.style={anchor=south,yshift=0.6cm},
        roundlabel/.style={anchor=east,xshift=-0.6cm},
        font=\scriptsize,
    ]
        \begin{scope}[yshift=0cm]
            \node [execlabel] at (0,0) {$G_1$};

            \node [leader, honact] (L) at (4,0) {$\ell$};

            \def\t{0}
            \node [roundlabel] at (0,-\t) {$t=\t$};
            \node [node, big, honslp] (A\t) at (0,-\t) {$A$};
            \node [node, sml, adv] (B\t) at (1,-\t) {$B$};
            \node [node, big, honslp] (C\t) at (2,-\t) {$C$};
            \node [node, big, honact] (S\t) at (3,-\t) {$S$};

            \def\t{1}
            \node [roundlabel] at (0,-\t) {$t=\t$};
            \node [node, big, honact] (A\t) at (0,-\t) {$A$};
            \node [node, sml, adv] (B\t) at (1,-\t) {$B$};
            \node [node, big, honslp] (C\t) at (2,-\t) {$C$};
            \node [node, big, honslp] (S\t) at (3,-\t) {$S$};

            \def\t{2}
            \node [roundlabel] at (0,-\t) {$t\geq\t$};
            \node [node, big, honact] (A\t) at (0,-\t) {$A$};
            \node [node, sml, adv] (B\t) at (1,-\t) {$B$};
            \node [node, big, honslp] (C\t) at (2,-\t) {$C$};
            \node [node, big, honslp] (S\t) at (3,-\t) {$S$};

            \draw [sim1] (L) to [simedge] node [pos=0.1,right] {1} (A1);
            \draw [sim1] (L) to [simedge] (B1);
            \draw [correct] (B0) to [simedge] (A1);
            \draw [omit] (B1) to [simedge] node [midway,above] {$\emptyset$} (A2);
        \end{scope}

        \begin{scope}[yshift=-5cm]
            \node [execlabel] at (0,0) {$G_2$};

            \node [leader, adv] (L) at (4,0) {$\ell$};

            \def\t{0}
            \node [roundlabel] at (0,-\t) {$t=\t$};
            \node [node, big, honslp] (A\t) at (0,-\t) {$A$};
            \node [node, sml, honact] (B\t) at (1,-\t) {$B$};
            \node [node, big, adv] (C\t) at (2,-\t) {$C$};
            \node [node, big, honact] (S\t) at (3,-\t) {$S$};

            \def\t{1}
            \node [roundlabel] at (0,-\t) {$t=\t$};
            \node [node, big, honact] (A\t) at (0,-\t) {$A$};
            \node [node, sml, honact] (B\t) at (1,-\t) {$B$};
            \node [node, big, adv] (C\t) at (2,-\t) {$C$};
            \node [node, big, honslp] (S\t) at (3,-\t) {$S$};

            \def\t{2}
            \node [roundlabel] at (0,-\t) {$t\geq\t$};
            \node [node, big, honact] (A\t) at (0,-\t) {$A$};
            \node [node, sml, honact] (B\t) at (1,-\t) {$B$};
            \node [node, big, adv] (C\t) at (2,-\t) {$C$};
            \node [node, big, honslp] (S\t) at (3,-\t) {$S$};

            \draw [sim1] (L) to [simedge] node [pos=0.1,left,yshift=2pt] {1} (A1);
            \draw [sim0] (L) to [simedge] node [pos=0.1,right] {0} (C1);
            \draw [sim1] (A1) to [simedge] ([xshift=-2pt]B2.north);
            \draw [sim0] (C1) to [simedge] ([xshift=+2pt]B2.north);
            \draw [omit] (B1) to [simedge] node [midway,above] {$\emptyset$} (A2);
        \end{scope}

        \begin{scope}[yshift=-5cm,xshift=6.5cm]
            \node [execlabel] at (0,0) {$G_3$};

            \node [leader, adv] (L) at (4,0) {$\ell$};

            \def\t{0}
            \node [roundlabel] at (0,-\t) {$t=\t$};
            \node [node, big, adv] (A\t) at (0,-\t) {$A$};
            \node [node, sml, honact] (B\t) at (1,-\t) {$B$};
            \node [node, big, honslp] (C\t) at (2,-\t) {$C$};
            \node [node, big, honact] (S\t) at (3,-\t) {$S$};

            \def\t{1}
            \node [roundlabel] at (0,-\t) {$t=\t$};
            \node [node, big, adv] (A\t) at (0,-\t) {$A$};
            \node [node, sml, honact] (B\t) at (1,-\t) {$B$};
            \node [node, big, honact] (C\t) at (2,-\t) {$C$};
            \node [node, big, honslp] (S\t) at (3,-\t) {$S$};

            \def\t{2}
            \node [roundlabel] at (0,-\t) {$t\geq\t$};
            \node [node, big, adv] (A\t) at (0,-\t) {$A$};
            \node [node, sml, honact] (B\t) at (1,-\t) {$B$};
            \node [node, big, honact] (C\t) at (2,-\t) {$C$};
            \node [node, big, honslp] (S\t) at (3,-\t) {$S$};

            \draw [sim1] (L) to [simedge] node [pos=0.1,left,yshift=2pt] {1} (A1);
            \draw [sim0] (L) to [simedge] node [pos=0.1,right] {0} (C1);
            \draw [sim1] (A1) to [simedge] ([xshift=-2pt]B2.north);
            \draw [sim0] (C1) to [simedge] ([xshift=+2pt]B2.north);
            \draw [omit] (B1) to [simedge] node [midway,above] {$\emptyset$} (C2);
        \end{scope}

        \begin{scope}[yshift=0cm,xshift=6.5cm]
            \node [execlabel] at (0,0) {$G_4$};

            \node [leader, honact] (L) at (4,0) {$\ell$};

            \def\t{0}
            \node [roundlabel] at (0,-\t) {$t=\t$};
            \node [node, big, honslp] (A\t) at (0,-\t) {$A$};
            \node [node, sml, adv] (B\t) at (1,-\t) {$B$};
            \node [node, big, honslp] (C\t) at (2,-\t) {$C$};
            \node [node, big, honact] (S\t) at (3,-\t) {$S$};

            \def\t{1}
            \node [roundlabel] at (0,-\t) {$t=\t$};
            \node [node, big, honslp] (A\t) at (0,-\t) {$A$};
            \node [node, sml, adv] (B\t) at (1,-\t) {$B$};
            \node [node, big, honact] (C\t) at (2,-\t) {$C$};
            \node [node, big, honslp] (S\t) at (3,-\t) {$S$};

            \def\t{2}
            \node [roundlabel] at (0,-\t) {$t\geq\t$};
            \node [node, big, honslp] (A\t) at (0,-\t) {$A$};
            \node [node, sml, adv] (B\t) at (1,-\t) {$B$};
            \node [node, big, honact] (C\t) at (2,-\t) {$C$};
            \node [node, big, honslp] (S\t) at (3,-\t) {$S$};

            \draw [sim0] (L) to [simedge,in distance=1.5em] node [pos=0.1,right] {0} ([xshift=+2pt]C1.north);
            \draw [sim0] (L) to [simedge] (B1);
            \draw [correct] (B0) to [simedge] ([xshift=-2pt]C1.north);
            \draw [omit] (B1) to [simedge] node [midway,above] {$\emptyset$} (C2);
        \end{scope}

    \end{tikzpicture}
    \caption{%
        Illustration of the four executions considered in the proof of \cref{thrm:good-case-BB-3-rounds}. 
        Green labels and arrows indicate correct behavior, or emulation of it, with input $1$. 
        Blue labels and arrows indicate correct behavior, or emulation of it, with input $0$. 
        Corrupt parties are colored red, correct parties orange.
        Translucent sets indicate that the set of parties is not awake at that round.
        The behavior of correct parties is well defined from the protocol description $\Pi$, so we omit most of the corresponding arrows to maintain simplicity. 
        Lack of an arrow from a set of corrupt parties to a set of correct parties indicates omission of all corresponding messages.
        Dashed red arrows labeled ``$\emptyset$'' indicate messages corresponding to receiving nothing from the leader $\ell$ in round $t=0$.
        Black arrows indicate correct round $t=0$ behavior.%
    }
    \label{fig:bb-lower-bound}
\end{figure}

\begin{proof}[Proof of \cref{thrm:good-case-BB-3-rounds}]
    Let $n$ be a natural number and let $\rho'\in [0,1]$ such that $\rho'>\rho\coloneqq1-\frac{1}{\varphi}$. In particular, this implies that $\rho'>(1-\rho)^2$, as $1-\frac{1}{\varphi}$ is a solution of the equation $x=(1-x)^2$. Assume towards a contradiction that there exists a $\rho'$-secure $\BB$ protocol $\Pi$ under dynamic participation with good-case latency $2$.
    Let $\ell$ be the designated leader. We consider some natural number $n$, and we define the following disjoint sets of parties $A,B,C$ such that $|A|=|C|=\lfloor\rho n\rfloor$, $|B|=\lceil\rho(1-\rho)n\rceil$. 
    Note that $\rho n+\rho n+\rho(1-\rho)n=(1-\rho)^2n +\rho n+\rho(1-\rho)n=n$. As such, since none of these added expressions are integers, we have that $|A|+|B|+|C|+|\set{\ell}|=n$. We further consider an additional disjoint set $S$ of parties where $|S|=|A|=|C|$.

    We now define and describe four executions of $\Pi$, which we denote by $G_1,G_2,G_3,G_4$. 
    See \cref{fig:bb-lower-bound} for an illustration.
    \begin{itemize}
        \item $G_1$: In round $t=0$, the parties in $S,B$ and $\ell$ are awake. The parties in $S$ are correct. In particular, the parties in $C$ are not active at all throughout the protocol. Furthermore, the parties in $B$ are corrupt, and are thus awake at all rounds. The leader $\ell$ is correct with input $1$. In round $t=1$, the parties in $S$ fall asleep, and the parties in $A$ wake up and remain awake for the remainder of the execution. The parties in $A$ are correct, and the parties in $B$ are corrupt. Note that up to an additive factor of $1$ due to the ceiling and floor, one has that $|A|+|B|=\rho n +\rho(1-\rho)n=\rho(2-\rho) n= (1-(1-\rho)^2)n=(1-\rho) n$, and thus for a sufficiently large $n$ \footnote{Namely, $n$ should be large enough so that ceil in $|B|$ does not violate the corruption bound $\rho'$, which is bounded away from $\rho$ by $\rho'-\rho>0$.}, it holds that
        $|B|=\lceil(1-\rho)\rho n\rceil<\rho'(|A|+|B|) $, and so the adversary is $\rho'$-bounded. The parties in $B$ behave correctly in round $t=0$, and in round $t=1$ they behave as if they have not received the round one messages of $\ell$. 
        \item $G_2$: In round $t=0$, the parties in $S,B,C$ and $\ell$ are awake. $S,B$ are correct. The leader $\ell$ is corrupt and the parties in $C$ are corrupt, and are thus awake throughout the whole execution. In round $t=1$, $S$ fall asleep, and the parties in $A$ wake up and remain awake for the remainder of the execution. All parties in $A,B$ are correct. Note that for a sufficiently large $n$, the adversary is $\rho'$-bounded as $|C|+|\set{\ell}|=\lfloor \rho n\rfloor +1$ by definition and there are $n$ awake parties. The leader $\ell$ behaves to $A$ as if it was correct with input $1$ in rounds $t=0$ and $t=1$ and then halts; it does not send any messages to the parties in $B$. The parties in $C$ do not send any round $t=0$ messages, and do not send round $t=1$ messages to $A$; they behave to $B$ in round $t=1$ as if they received correct round $t=0$ messages from $\ell$ with input $0$. Note that this is doable as $\ell$ is corrupt. They behave correctly for the remainder of the execution, except for pretending to have not received round $t=1$ messages from $A$. 
        \item $G_3$: This is the dual execution to $G_2$. In round $t=0$, the parties in $S,B,A$ and $\ell$ are awake. $S,B$ are correct. The leader $\ell$ is corrupt and the parties in $A$ are corrupt, and are thus awake throughout the whole execution. In round $t=1$, $S$ fall asleep, and the parties in $C$ wake up and remain awake for the remainder of the execution. All parties in $C$ are correct. Note that the adversary is $\rho'$-bounded by the same argument as in $G_2$. The leader behaves to $C$ as if it was correct with input $0$ in rounds $t=0$ and $t=1$ and then halts; it does not send any messages to parties in $B$. The parties in $A$ do not send any round $t=0$ messages, and do not send round $t=1$ messages to $C$; they behave to $B$ in round $t=1$ as if they received correct round $t=0$ messages from $\ell$ with input $1$. Note that this is doable as $\ell$ is corrupt. They behave correctly for the remainder of the execution, except for pretending to have not received round $t=1$ messages from $C$. 
        \item $G_4$: This is the dual execution to $G_1$. In round $t=0$, the parties in $S,B$ and $\ell$ are awake. The parties in $S$ are correct. In particular, the parties in $A$ are not active at all throughout the execution. Furthermore, the parties in $B$ are corrupt, and are thus awake at all times. The leader $\ell$ is correct with input $0$. In round $t=1$, the parties in $S$ fall asleep, and the parties in $C$ wake up. The parties in $C$ are correct. Note that the adversary is $\rho'$-bounded by the same argument as in $G_1$. The parties in $B$ behave correctly in round $t=0$, and in round $t=1$ they behave as if they have not received the round $t=0$ messages of $\ell$.
    \end{itemize}

    With $G_1,G_2,G_3,G_4$ defined, we can now proceed with the following observations.
    \begin{itemize}
        \item The parties in $A$ decide $1$ in $G_1$ after 2 rounds. This is a consequence of the assumed good-case latency of $2$ property of $\Pi$, the leader being correct with input $1$ in execution $G_1$, and the adversary being $\rho'$-bounded.
        \item The parties in $C$ decide $0$ in $G_4$ after 2 rounds. This is a consequence of the assumed good-case latency of $2$ property of $\Pi$, the leader being correct with input $0$ in execution $G_4$, and the adversary being $\rho'$-bounded.
        \item In the first two rounds, the parties in $A$ \emph{cannot distinguish} between $G_1$ and $G_2$. In both executions, they receive the same set of messages from the exact same parties in the first two rounds. Thus, the parties in $A$ decide $1$ in $G_2$ after 2 rounds.
        \item In the first two rounds, the parties in $C$ \emph{cannot distinguish} between $G_3$ and $G_4$. In both executions, they receive the same set of messages from the exact same parties in the first two rounds. Thus, the parties in $C$ decide $0$ in $G_3$ after 2 rounds.
        \item The parties in $B$ \emph{cannot distinguish} between $G_2,G_3$. In $G_2$, the parties in $C$ are corrupt and falsely claim to not have received round $2$ messages from $A$, and in $G_3$, the parties in $A$ are corrupt and falsely claim to not have received round $2$ messages from $C$. In both $G_2,G_3$, the parties in $A,C$ behave correctly after round $2$. Thus, throughout both executions, $B$ sees the exact same set of messages, from the exact same sets of parties, with the same delivery timings. Thus, by termination, eventually the parties in $B$ decide the same value in both $G_2,G_3$.
    \end{itemize}
    Putting these observations together, we deduce, by the agreement property of $\Pi$, that $B$ outputs $1$ in $G_2$, and $0$ in $G_3$. This contradicts the last observation above, thus concluding the proof. 
\end{proof}

\subsection{BA Lower Bound}
\label{ssec:BA_lower_bound}
In this section, we prove \cref{thrm:good-case-BA-2-rounds}.

\begin{figure}[tbp]
    \centering
    \begin{tikzpicture}[
        x=1.1cm,
        node/.style={},
        big/.style={minimum width=0.8cm, minimum height=0.8cm, rounded corners=3pt},
        sml/.style={minimum width=0.6cm, minimum height=0.6cm, rounded corners=2pt},
        honact/.style={fill=YellowOrange},
        honslp/.style={fill=YellowOrange!20},
        adv/.style={fill=OrangeRed},
        input0/.style={color=RoyalBlue,anchor=base,yshift=1cm},
        input1/.style={color=Green,anchor=base,yshift=1cm},
        sim0/.style={-latex,color=RoyalBlue},
        sim1/.style={-latex,color=Green},
        omit/.style={-latex,dashed,color=Red},
        execlabel/.style={anchor=east,xshift=-0.8cm},
    ]
        \begin{scope}[yshift=0cm]
            \node [execlabel] at (0,0) {$G_1$};

            \node [node, big, honact] (A) at (0,0) {$A$};
            \node [node, sml, honact] (B) at (1,0) {$B$};
            \node [node, sml, adv] (C) at (2,0) {$C$};
            \node [node, big, honslp] (D) at (3,0) {$D$};

            \node [input1] at (0,0) {1};
            \node [input1] at (1,0) {1};
            \node [input0] at (2,0) {0};
        \end{scope}
        
        \begin{scope}[yshift=-3cm]
            \node [execlabel] at (0,0) {$G_2$};

            \node [node, big, honact] (A) at (0,0) {$A$};
            \node [node, sml, honact] (B) at (1,0) {$B$};
            \node [node, sml, honact] (C) at (2,0) {$C$};
            \node [node, big, adv] (D) at (3,0) {$D$};

            \node [input1] at (0,0) {1};
            \node [input1] at (1,0) {1};
            \node [input0] at (2,0) {0};

            \draw [sim0] (D.north) to [bend right=30] (C.north);
            \draw [sim0] (D.north) to [bend right=30] (B.north);
            \draw [omit] (D.south) to [bend left=20] node [midway,above] {Omit} (A.south);
        \end{scope}
        
        \begin{scope}[yshift=-3cm,xshift=6.5cm]
            \node [execlabel] at (0,0) {$G_3$};

            \node [node, big, adv] (A) at (0,0) {$A$};
            \node [node, sml, honact] (B) at (1,0) {$B$};
            \node [node, sml, honact] (C) at (2,0) {$C$};
            \node [node, big, honact] (D) at (3,0) {$D$};

            \node [input1] at (1,0) {1};
            \node [input0] at (2,0) {0};
            \node [input0] at (3,0) {0};

            \draw [sim1] (A.north) to [bend left=30] (B.north);
            \draw [sim1] (A.north) to [bend left=30] (C.north);
            \draw [omit] (A.south) to [bend right=20] node [midway,above] {Omit} (D.south);
        \end{scope}
        
        \begin{scope}[yshift=0cm,xshift=6.5cm]
            \node [execlabel] at (0,0) {$G_4$};

            \node [node, big, honslp] (A) at (0,0) {$A$};
            \node [node, sml, adv] (B) at (1,0) {$B$};
            \node [node, sml, honact] (C) at (2,0) {$C$};
            \node [node, big, honact] (D) at (3,0) {$D$};

            \node [input1] at (1,0) {1};
            \node [input0] at (2,0) {0};
            \node [input0] at (3,0) {0};
        \end{scope}

    \end{tikzpicture}
    \caption{%
    Illustration of the four executions considered in the proof of \cref{thrm:good-case-BA-2-rounds}. 
    Green labels and arrows indicate correct behavior, or emulation of it, with input $1$. 
    Blue labels and arrows indicate correct behavior, or emulation of it, with input $0$. 
    Corrupt parties in each execution are colored red, correct parties orange.
    Translucent sets indicate that the set of parties is not participating in the given execution (inactive).
    Dashed red arrows labeled ``Omit'' indicate that no messages are sent from the source set of the arrow to its target set in round $t=0$.%
    }
    \label{fig:ba-lower-bound}
\end{figure}
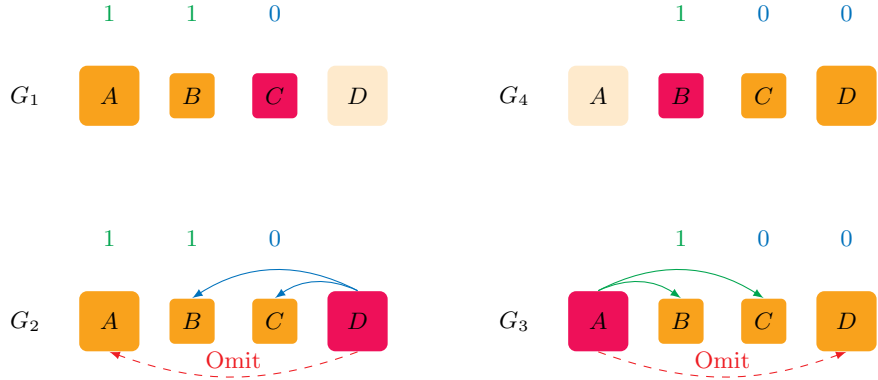

\begin{proof}[Proof of \cref{thrm:good-case-BA-2-rounds}]
    Let $n$ be a natural number and let $\rho'\in [0,1]$ such that $\rho'>\rho\coloneqq1-\frac{1}{\sqrt{2}}$. In particular, this implies that $\rho'>(1-\rho)^2-\rho(1-\rho)$, as $1-\frac{1}{\sqrt{2}}$ is a solution for $x=(1-x)^2-x(1-x)$. Assume towards a contradiction that there exists a $\rho'$-secure $\BA$ protocol $\Pi$ with good-case latency $1$ under unknown participation. We consider a set of $n$ parties, and a partition of them into disjoint sets 
    $A,B,C,D$ such that $|A|=|D|=\lceil \rho n\rceil$, and $|B|=|C|=\lfloor\rho(1-\rho)n\rfloor$. Note that by substituting $\rho$ with $(1-\rho)^2-\rho(1-\rho)$, one gets $\rho n+\rho(1-\rho) n+(1-\rho)^2 n=n$. Since none of these added expressions are integer, and by the fact that $A,D$ are with ceil while $B,C$ are with floor, we get that $|A|+|B|+|C|+|D|=n$. Hence this is a valid partition of the $n$ parties. We now define and describe four executions of $\Pi$, which we denote by $G_1,G_2,G_3,G_4$. 
    See \cref{fig:ba-lower-bound} for an illustration.

    \begin{itemize}
        \item $G_1$: The participating parties are $A,B,C$. The parties in $A,B$ are correct and $C$ is corrupt. Note that up to an additive factor of $1$, we have that $|A\cup B|=|A|+|B|=(1-\rho)^2 n$, and 
        $|A|+|B|+|C|=\rho(1-\rho)n+(1-\rho)^2n=(1-\rho) n$, thus for a sufficiently large $n$, we have that $|C|\leq \rho|A\cup B\cup C|<\rho'|A\cup B\cup C|$, which means the adversary is $\rho'$-bounded. The parties in $A,B$ have input $1$, and the parties in $C$ behave as if they are correct with input $0$.
        \item $G_2$: The participating parties are $A,B,C,D$. The parties in $A,B,C$ are correct, and the parties in $D$ are corrupt. Note that $|A\cup B\cup C\cup D|=n$, and that $|D|=\lceil \rho n\rceil $ as $\rho=(1-\rho)^2-\rho(1-\rho)$. Thus, for a sufficiently large $n$, one has that $\lceil \rho n\rceil<\rho' n$, and so the adversary is $\rho'$-bounded. The parties in $A,B$ have input $1$, the parties in $C$ have input $0$. The parties in $D$ behave as if they are correct with input $0$, \emph{except for} not sending any round $t=0$ messages to $A$, and pretending to have not received round $t=0$ messages from $A$. They behave correctly in all the remaining rounds. 
        \item $G_3$: This is the dual execution to $G_2$. The participating parties are $A,B,C,D$. The parties in $C,B,D$ are correct, and the parties in $A$ are corrupt. Note that the adversary is $\rho'$-bounded by the same argument as in $G_2$. The parties in $C,D$ have input $0$, the parties in $B$ have input $1$. The parties in $A$ behave as if they are correct with input $1$, \emph{except for} not sending any round $t=0$ messages to $D$, and pretending to have not received round $t=0$ messages from $D$. They behave correctly in all the remaining rounds. 
        \item $G_4$: This is the dual execution to $G_1$. The participating parties are $B,C,D$. The parties in $C,D$ are correct, and the parties in $B$ are corrupt. Note that the adversary is $\rho'$-bounded by the same argument as in $G_1$. The parties in $C,D$ have input $0$, and the parties in $B$ behave as if they are correct with input $1$.
    \end{itemize}

    With $G_1,G_2,G_3,G_4$ defined, we can now proceed with the following observations.
    \begin{itemize}
        \item All correct parties in $G_1$ have the same input $1$. Thus, by the validity of $\BA$ and the good-case latency of $1$ property of $\Pi$, the parties in $A$ decide $1$ at the end of the first round.
        \item All correct parties in $G_4$ have the same input $0$. Thus, by the validity of $\BA$ and the good-case latency of $1$ property of $\Pi$, the parties in $D$ decide $0$ at the end of the first round.
        \item At the end of the first round, the parties in $A$ cannot distinguish between $G_1,G_2$. Thus, also in $G_2$, the parties in $A$ decide $1$ at the end of the first round. By the agreement property of $\BA$, we thus have that both $B,C$ eventually decide $1$ in $G_2$.
        \item At the end of the first round, the parties in $D$ cannot distinguish between $G_3,G_4$. Thus, also in $G_3$, the parties in $D$ decide $0$ at the end of the first round. By the agreement property of $\BA$, we thus have that both $B,C$ eventually decide $0$ in $G_3$.
        \item The parties in $B,C$ cannot distinguish between $G_2,G_3$, as in both executions they receive the exact same sets of messages, from the same sets of parties, with the same delivery timings, as both the parties in $A$ and $D$ claim to have not heard from the other in round $t=1$. Thus, by termination, eventually the parties in $B,C$ decide \emph{the same} value in both $G_2,G_3$.
    \end{itemize}

    From the combination of the last three observations, we arrive at a contradiction, as the parties in $B,C$ output $1$ in $G_2$, $0$ in $G_3$, but also the same value in both $G_2,G_3$. This concludes the proof.
\end{proof}

\section{$\BB$ and $\BA$ with Optimal Good-Case Latency}
\label{sec:fast_path_BB_BA}

In this section, we prove the following theorems.
\begin{theorem}\label{thrm:BB_fast_path}
    For all $\rho \leq 1 - \frac{1}{\varphi} \approx 0.382$
    where $\varphi = \frac{1+\sqrt{5}}{2} \approx 1.618$, 
    there exists a $\rho$-secure protocol (namely $\BBFP$ below) solving $\BB$ under dynamic participation with good-case latency $\Tgc=2$. 
\end{theorem}

\begin{theorem}\label{thrm:BA_fast_path}
    For all $\rho \leq 1-\frac{1}{\sqrt{2}} \approx 0.293$,
    there exists a $\rho$-secure protocol (namely $\BAFP$ below) solving $\BA$ under dynamic participation with good-case latency $\Tgc=1$.
\end{theorem}

\begin{theorem}\label{thrm:BB_fast_path_unknown_participation}
    For all $\rho \leq 1/2$,
    there exists a $\rho$-secure protocol (namely $\BBUP$ below) solving $\BB$ under unknown participation with good-case latency $\Tgc=2$.
\end{theorem}

\subsection{$\BB$ with Good-Case Latency $\Tgc=2$}
\label{ssec:BB_fast_path}

In this section, we prove \cref{thrm:BB_fast_path}. 
In this protocol, we make black-box use of the $\BA$ protocol from~\cite{DACS}, which we denote by $\Pi_\BA$, and which is $\frac{1}{2}$-secure under dynamic participation.

\begin{protocol}{$\BBFP$}
    \label{prot:BB_fast_path}
    \begin{enumerate}
        \item[] \textbf{Setup.} Leader $\ell$ with input value $b_\ell$. Instructions for party $p$. 
        \item \textbf{Round $t=0$}: $\ell$ multicasts $\ip{\echo,b_\ell}_\ell$.
        \item \textbf{Round $t=1$}: If received $\ip{\echo,u}_\ell$ for some $u$, then multicast $\ip{\vote,u}_p$ along with $\ell$'s $\echo$ message. Else, multicast $\ip{\vote,\bot}$.
        \item \textbf{Round $t=2$}: Let $V$ denote the set of parties from which $p$ received a valid $\vote$ message. Denote by $V_u$ the set of parties from which $p$ received a valid $\vote$ for value $u$. Let $u$ be the value $\arg\max\limits_{u\in I} |V_u|$, breaking ties arbitrarily. 
        Multicast $\ip{\forward,V_u}_p$, along with the corresponding valid $\vote$ messages for $u$ from the parties in $V_u$.
        
        \textbf{Early deciding condition:} If the following holds:
        \begin{enumerate}
            \item Did not detect equivocation from the leader $\ell$. I.e., did not receive two $\echo$ (either forwarded or directly from $\ell$) messages signed by the leader $\ell$ for values $u\neq u'$.
            \item It holds that $|V_u|> (1-\rho)|V|$.
        \end{enumerate}
         Then decide $u$ and multicast $\ip{\decided,u}_p$.

         \item \textbf{Round $t\geq 3$:}
         \begin{enumerate}
             \item \textbf{Early deciding:} For a value $u'$, denote by $D_{u'}$ the set of parties from which $p$ received a $\decided$ message for $u'$, and by $D^*$ the set of parties from which $p$ received a $\forward$ message. If there exists a value $u$ such that $|D_u|>(1-\rho) |D^*|$, then decide $u$. 
             \item Let $u'$ be the value for which $p$ received the largest $V_u$ set among all received $\forward$ messages. Run $\Pi_\BA$ as instructed that commences in round $t=3$ with input $u'$, and decide the output of $\Pi_\BA$, if did not decide before. 
         \end{enumerate}

    \end{enumerate}
\end{protocol}

\begin{proof}[Proof of \cref{thrm:BB_fast_path}]
Let $\rho$ be as in the premise of \cref{thrm:BB_fast_path}.
Our protocol is given in $\BBFP$. We now analyze its security and efficiency. Termination follows from the termination of $\Pi_\BA$, as all parties decide and halt once $\Pi_\BA$ terminates. We next handle agreement.
\begin{claim}\label{claim:BB_fast_path_agreement}
    $\BBFP$ satisfies agreement against any $\rho$-bounded adversary.
\end{claim}
\begin{claimproof}
    There are two cases: Either there exists a correct party that decides from the early deciding condition, i.e., prior to participating in $\Pi_\BA$, or not. Agreement is straightforward in the latter case, as no correct party multicasts a $\decided$ message, and thus no correct party decides due to an early deciding condition in all rounds $t\geq 2$.  Thus, due to $\Pi_\BA$ being $\frac{1}{2}$-secure and all correct parties deciding according to its output, agreement holds. 
    
    In the former case, let $p$ be a correct party deciding a value $u$ by the early deciding condition. 
    Denote by $V^p$ the set of parties from which $p$ received a valid $\vote$ message. We first argue that there cannot be two correct parties $p,q$, both awake at round $t=2$, that decide by the early deciding condition different values $u\neq u'$. If this was the case, then $p$ has $|V^p_u|>(1-\rho)|V^p|$, and $|V^q_{u'}|>(1-\rho)|V^q|$. As the network is synchronous and $\rho<\frac{1}{2}$, this in particular implies that $p$ has received at least one valid $\vote$ message for $u$ from a correct party, and $q$ received at least one valid $\vote$ message for $u'$ from a correct party. This implies, however, that both $p,q$ saw an equivocation from the leader $\ell$ at round $t=2$, and thus none of them would have triggered the early deciding condition. This in particular implies that there exists at most one value, for which any correct party sends a $\decided$ message for, and thus all correct parties deciding due to the early deciding condition decide the same value. 
    
    It remains to show that if some correct party $p$ decides a value $u$ via the early deciding condition at round $t=2$, then all parties eventually decide $u$. 
    Denoting the correct parties awake at round $t=1$ by $H$, we have that $H\subseteq V^p$. Furthermore, we have that $|V^p_u|>(1-\rho)|V^p|$, thus $|V^p_u|>(1-\rho)|H|$. Furthermore, $p$ detected no equivocation from $\ell$, hence no party in $H$ multicast a $\vote$ message for a value $u'\neq u$. Denoting by $F$ the number of corrupt parties, we have that $|F|<\rho(|H|+|F|)$, thus $|F|<\frac{\rho}{1-\rho} |H|$. In particular, we can deduce that any other correct party $q$ receives at most $|F|<\frac{\rho}{1-\rho}|H|$ valid $\vote$ messages for values different from $u$. By the behavior of the protocol, $p$ multicasts in round $t=2$ the message $\ip{\forward,V^p_u}_p$, and so all correct parties awake at rounds $\geq 3$ observe at least $(1-\rho)|H|$ vote messages in the $\forward$ message of $p$. Finally, note that $(1-\rho)|H|-|F|>(1-\rho)|H|-\frac{\rho}{1-\rho}|H|=\frac{(1-\rho)^2-\rho}{(1-\rho)}|H|>0$. The last inequality holds by the assumption that $\rho<(1-\rho)^2$. Thus, all correct parties awake at round $3$ commence $\Pi_\BA$ with input $u$. By the validity guarantee of $\Pi_\BA$, we thus have that all parties decide $u$, as required.
\end{claimproof}

Finally, we handle validity and good-case latency.
\begin{claim}\label{claim:BB_fast_path_validity}
    $\BBFP$ satisfies validity against any $\rho$-bounded adversary. Furthermore, $\BBFP$ has good-case latency $2$.
\end{claim}
\begin{claimproof}
    Assume $\ell$ is correct with input $u$. Then by the behavior of $\BBFP$, it multicasts $\ip{\echo,u}_\ell$ to all correct parties at round $t=0$. Then, at round $t=1$, all correct parties multicast $\ip{\vote,u}$ along with $\ell$'s $\echo$ message. Note that since $\ell$ is correct, no other valid $\vote$ for any other value other than $u$ can be sent, by correct or corrupt parties. Thus, at round $t=2$, no correct party detects equivocation from the leader, and all correct parties awake at $t=2$ see a valid $\vote$ message for $u$ from all correct parties awake at $t=1$. As the adversary is $\rho$-bounded, we thus have that $|V_u^p|>(1-\rho)|V^p|$ for all correct parties $p$ awake at $t=2$, and they all decide $u$ at $t=2$ and send corresponding $\decided$ messages. The same argument applied to the early deciding condition in rounds $t\geq 3$ then implies that any correct party $p$ decides $u$ in the first round $t\geq2$ in which it is awake, as required.
\end{claimproof}

    This concludes the proof of \cref{thrm:BB_fast_path}.
\end{proof}

\subsection{$\BA$ with Good-Case Latency $\Tgc=1$}
\label{ssec:BA_fast_path}

In this section, we prove \cref{thrm:BA_fast_path}.
In this protocol, we make black-box use of the $\BA$ protocol from~\cite{DACS}, which we denote by $\Pi_\BA$, and which is $\frac{1}{2}$-secure under dynamic participation.

\begin{protocol}{\BAFP}
\label{prot:BA_fast_path}
    \begin{enumerate}
        \item[] \textbf{Setup.} Each party $p$ has input value $u_p$. Instructions for party $p$.
        \item \textbf{Round $t=0$}: Multicast $\ip{\echo, u_p}_p$.
        \item \textbf{Round $t=1$}: Denote by $E$ the set of parties from which $p$ received an $\echo$ message. Let $u$ be the value $\arg\max\limits_{u\in I} |E_u|$, breaking ties arbitrarily. Multicast $\ip{\forward,E_u}_p$, along with the corresponding $\echo$ messages for $u$.
        
        \textbf{Early deciding condition:} If $|E_u|>(1-\rho)|E|$, then decide $u$, and multicast $\ip{\decided,u}_p$.
        \item \textbf{Round $t\geq 2$:}
            \begin{enumerate}
                \item \textbf{Early Deciding:} For a value $u'$, denote by $D_{u'}$ the set of parties from which $p$ received a $\decided$ message for $u'$, and by $D^*$ the set of parties from which $p$ received a $\forward$ message. If there exists a value $u$ such that $|D_u|>(1-\rho) |D^*|$, then decide $u$. 
                \item Let $u'$ be the value for which $p$ received the largest $E_u$ set among all received $\forward$ messages. Run $\Pi_\BA$ as instructed that commences in round $t=2$ with input $u'$, and decide the output of $\Pi_\BA$, if did not decide before.
            \end{enumerate}
    \end{enumerate}
\end{protocol}

\begin{proof}[Proof of \cref{thrm:BA_fast_path}]
    Let $\rho$ be as in the premise of \cref{thrm:BA_fast_path}. Our protocol is given in $\BAFP$. We now analyze its security and efficiency. Termination follows from the termination of $\Pi_\BA$, as all correct parties decide and halt after $\Pi_\BA$ outputs. Next, we handle agreement. 
    \begin{claim}\label{claim:BA_fast_path_agreement}
        $\BAFP$ satisfies agreement against any $\rho$-bounded adversary.
    \end{claim}
    \begin{claimproof}
        There are two cases: either there exists a correct party that decides due to the early deciding condition at round $t=1$, or not. The proof can be quickly concluded in the latter case, as then no correct party multicasts a $\decided$ message, and thus no correct party decides according to the early deciding condition in rounds $t\geq 2$. Thus all correct parties decide according to the output of $\Pi_\BA$, which guarantees agreement. 
        
        We are now left with the case that there exists a correct party that decides a value $u$ according to the early deciding condition at round $t=1$. First we prove that this can occur for at most a single value. Assume that there exist two correct parties $p,q$ that decide according to the early deciding condition in round $t=1$ values $u,u'$, respectively, where $u\neq u'$. Denote by $E^p,E^q$ the set of $\echo$ messages received by $p,q$, respectively. This implies that $|E^p_u|>(1-\rho)|E^p|$, and $|E^q_{u'}|>(1-\rho)|E^q|$. Denoting by $H$ the set of correct parties awake at round $t=0$, we have that $H\subseteq E^p$ and $H\subseteq E^q$, so $|E^p_u|>(1-\rho)|H|$, and $|E^q_{u'}|>(1-\rho)|H|$. 
        Denoting by $F^p,F^q$ the number of corrupt parties that sent an $\echo$ message to $p,q$, respectively. We have that at least $|E^p_u|-|F^p|$ of the $\echo$ messages for $u$ received by $p$ came from correct parties, and similarly at least $|E^q_{u'}|-|F^q|$ of the $\echo$ messages for $u'$ received by $q$ came from correct parties. We thus have that: $$|E^p_u|-|F^p|>(1-\rho)(|H|+|F^p|)-|F^p|=(1-\rho)|H|-\rho|F^p|,$$ and similarly: $$|E^q_{u'}|-|F^q|>(1-\rho)(|H|+|F^q|)-|F^q|=(1-\rho)|H|-\rho|F^q|.$$ As $F^q,F^p\subseteq F$, and $|F|<\frac{\rho}{1-\rho}|H|,$ we have: 
        \begin{gather*}
        |E^p_u|-|F^p|>(1-\rho)|H|-\frac{\rho^2}{1-\rho}|H|=\frac{(1-\rho)^2-\rho^2}{1-\rho}|H|=\frac{1-2\rho}{1-\rho}|H|\\=(1-\frac{\rho}{1-\rho})|H|>(1-(1-2\rho))|H|=2\rho |H|.
        \end{gather*}
        The last inequality holds due to the assumption $\rho<(1-\rho)(1-2\rho)$.        
        This allows us to deduce that less than $(1-2\rho)|H|$ correct parties sent an $\echo$ message for a value other than $u$. Thus, $(1-2\rho)|H|\geq |E^q_{u'}|-|F^q|>(1-\rho)|H|-\rho|F^q|$ which implies $|F^q|>|H|$, a contradiction. Thus, there exists at most a single value $u$ for which some correct party decides $u$ via the early deciding condition in round $t=1$. 
        
        This implies that correct parties can only send $\decided$ messages for a single value $u$, and so any correct party deciding according to the early decision condition in round $t\geq 2$ decides $u$ as well, and agreement is maintained. 

        It remains to show that when a correct party $p$ decides $u$ due to the early deciding condition in round $t=1$, all correct parties that decide on the $\Pi_\BA$ output also decide $u$.
        By the early deciding condition in round 1,  $|E^p_u|>(1-\rho)(|H|+|F^p|)$, and at least $|E^p_u|-|F^p|>(1-\rho)|H|-\rho|F^p|$ of those $\echo$ messages for $u$ came from correct parties. Thus, less than $\rho|H|+\rho |F^p|$ correct parties multicast an $\echo$ message for any value $u'\neq u$. This implies that the size of the $E_{u'}$ set in any $\forward$ message received by any correct party for a value $u'\neq u$ is bounded by 
        \begin{gather*}
        \rho|H|+\rho |F^p|+|F|<\rho(|H|+|F^p|)+\frac{\rho}{1-\rho}|H|\\\overset{(1)}{<}\rho|H|+\rho|F^p|+(1-2\rho)|H|<(1-\rho)(|H|+|F^p|)<|E^p_u|.
        \end{gather*}
        Where (1) holds due to the assumption $\rho<(1-\rho)(1-2\rho)$.
        Thus, the $\forward$ message sent by $p$ with the set $E^p_u$ is the largest set of $\echo$ messages witnessed by any correct party in round $t=2$, thus by the behavior of $\BAFP$, all correct parties awake at round $t=2$ commence $\Pi_\BA$ with input $u$. The validity property of $\Pi_\BA$ then ensures that all correct parties output $u$ from $\Pi_\BA$, thus all correct parties agree, as required.  
    \end{claimproof}

    Next, we handle validity and good-case latency.
    \begin{claim}\label{claim:BA_fast_path_validity}
        $\BAFP$ satisfies validity against any $\rho$-bounded adversary. In particular, $\BAFP$ has good-case latency of $1$.
    \end{claim}

    \begin{claimproof}
        Assume that all correct parties awake at round $t=0$ have the same input value $u$, then, due to the adversary being $\rho$-bounded, all correct parties awake at round $t=1$ observe $|E_u|>(1-\rho)|E|$, as all correct parties send an $\echo$ message for $u$. In particular, \emph{all} correct parties awake in round $t=1$ decide $u$ by the early deciding condition and multicast a $\decided$ message for $u$. An identical argument then implies that any correct party $p$, in the first round $t\geq 2$ in which it is awake, observes $|D_u|>(1-\rho)|D^*|$, and thus $p$ decides in the first round $t\geq 1$ in which it is awake. This proves that $\BAFP$ satisfies validity and has $\Tgc=1$.
    \end{claimproof}

    This concludes the proof of \cref{thrm:BA_fast_path}.
\end{proof}

\subsection{$\BB$ for Unknown Participation with Good-Case Latency $\Tgc=2$}
\label{ssec:BB_fast_path_unknown_participation}

In this section, we prove \cref{thrm:BB_fast_path_unknown_participation}.
In this protocol, we make black-box use of the $\BA$ protocol from~\cite{DACS}, which we denote by $\Pi_\BA$, and which is $\frac{1}{2}$-secure under dynamic participation.

\begin{protocol}{$\BBUP$}
    \label{prot:BB_fast_path_unknown_participation}
    \begin{enumerate}
        \item[] \textbf{Setup.} Leader $\ell$ with input value $b_\ell$. Instructions for party $p$.
        \item \textbf{Round $t=0$}: If $p=\ell$, multicast $\ip{\echo,b_\ell}_\ell$. Else, multicast $\ip{\echo,\bot}_p$.
        \item \textbf{Round $t=1$}: Denote by $E$ the set of parties from which $p$ received an $\echo$ message. Multicast $\ip{\forward,E}_p$, along with the corresponding $\echo$ messages. If $\ell\in E$ and received $\ip{\echo,u}_\ell$, multicast $\ip{\vote,u}_p$ along with $\ell$'s $\echo$.
        \item \textbf{Round $t=2$}: Let $G$ denote the set of parties from which $p$ received a $\forward$ message, and let $V$ denote the set of parties from which $p$  received a valid $\vote$ message. Denote by $E$ the set of parties from which $p$ received an $\echo$ message. Denote by $E^*\subseteq E$ the set of parties $p$ for which an $\echo$ message from $p$ appeared in more than $\frac{|G|}{2}$ of the received $\forward$ messages. Let $u$ be the value $\arg\max\limits_{u\in I} |V_u\cap E^*|$. Multicast $\ip{\echotwo,V_u\cap E^*}_p$, along with the corresponding valid $\vote$ messages. 

        \textbf{Early deciding condition:} If the following holds:
        \begin{enumerate}
            \item Did not detect equivocation from the leader $\ell$, i.e., did not receive two valid $\vote$ messages for values $u\neq u'$.
            \item It holds that $|V_u\cap E^*|>\frac{|E|}{2}$.
        \end{enumerate}

        Then decide $u$ and multicast $\ip{\decided,u}_p$.
        \item \textbf{Round $t\geq 3$:}
        \begin{enumerate}
            \item Let $u'$ be the value for which $p$ received the largest $V_u\cap E^*$ amongst all received $\echotwo$ messages. Run $\Pi_\BA$ as instructed that commences in round $t=3$ with input $u'$, and decide the output of $\Pi_\BA$, if did not decide before. 
        \end{enumerate}
    \end{enumerate}
\end{protocol}

\begin{proof}[Proof of \cref{thrm:BB_fast_path_unknown_participation}]
    Our protocol is given in $\BBUP$. We now analyze its security and efficiency. Termination follows from the termination of $\Pi_\BA$, as all parties decide and halt once $\Pi_\BA$ terminates. We next handle agreement. Denote by $n$ the total number of parties, which is unknown to the correct parties.
    \begin{claim}\label{claim:BB_unknown_participation_agreement}
        $\BBUP$ satisfies agreement against any $\frac{1}{2}$-bounded adversary.
    \end{claim}
    \begin{claimproof}
        There are two cases: either there exists a correct party that decides from the early deciding condition, i.e., prior to participating in $\Pi_\BA$, or not. Agreement is straightforward in the latter case, with $\Pi_\BA$ being $\frac{1}{2}$-secure and all correct parties deciding its output. We next focus on the former case.
        
        Denote by $V^p$ the set of parties from which $p$ received a valid $\vote$ message, and denote by $E^p,E^{*,p}$ the $E,E^*$ sets observed by party $p$ in round 2, respectively. We first argue that there cannot be two correct parties $p,q$ that decided due to the early deciding condition in round $t=2$ values $u\neq u'$, respectively. Suppose this was the case, then $p$ has observed $|V_u^p\cap E^{*,p}|>\frac{E^p}{2}$. Denoting by $H$ the set of correct parties, first observe that $H\subseteq E^{*,p}$, as a correct party sends an $\echo$ message to all parties, and all correct parties include that $\echo$ message in their $\forward$ messages. Since $H$ constitutes a majority of the parties, all correct parties observe the $\echo$ messages of correct parties as part of their $E^*$ set. This in particular implies that there exists a correct party in $V_u^p\cap E^{*,p}$, and the same argument yields that there exists a correct party in $V_{u'}^q\cap E^{*,q}$. This implies, however, that all correct parties observed in round $t=2$ a valid $\vote$ message for both $u,u'$, and thus no correct party decides by the early deciding condition in round $t=2$. This implies that there exists at most a single value $u$, such that a correct party decides $u$ due to the early deciding condition in round $t=2$. In particular, this also implies that if any correct party decides due to the early deciding condition at a round $t\geq 3$, it can only possibly decide $u$, as no $\decided$ message can be sent by a correct party for any other value. 
        
        Now suppose that a correct party $p$ decided $u$ due to the early deciding condition at round $t=2$. This implies that $|V_u^p\cap E^{*,p}|>\frac{|E^p|}{2}$. This also implies that no correct party multicast a valid $\vote$ message for any other value.
        We next observe that for any correct parties $p,q$, we have $E^{*,p} \subseteq E^q$, and vice versa.
        Consider any $z\in E^{*,p}$. The $\echo$ message of $z$ appeared in the majority of $\forward$ messages received by $p$, at least one of which was sent by a correct party, and thus all correct parties have seen the $\echo$ message of $z$. 
        For a correct party $q$, denote by $F^q$ the set of corrupt parties in $E^{*,q}$. We thus have $F^q\subseteq E^{*,q} \subseteq E^p$. We thus have that $|V_{u'}^q\cap E^{*,q}| \leq |F^q|<\frac{|H|+|F^q|}{2} \leq \frac{|E^p|}{2}<|V_u^p\cap E^{*,p}|$. Thus we have that the $\echotwo$ message from $p$ constitutes the largest set observed by $q$, when intersected with $E^{*,q}$. And so, in round $t=3$, $q$ commences $\Pi_\BA$ with input $u$. This holds for all correct parties, which triggers the validity property of $\Pi_\BA$, and thus all correct parties output $u$ from $\Pi_\BA$ and decide $u$, as required.
    \end{claimproof}

    Next, we handle validity and good-case latency. 
    \begin{claim}\label{claim:BB_unknown_participation_validity}
        $\BBUP$ satisfies validity against any $\frac{1}{2}$-bounded adversary. Furthermore, $\BBUP$ has good-case latency $2$.
    \end{claim}
    \begin{claimproof}
        Assume $\ell$ is correct with input $u$. Then by the behavior of $\BBUP$, it multicasts $\ip{\echo,u}_\ell$ to all correct parties at round $t=0$, and all other correct parties multicast $\ip{\echo,\bot}$ at round $t=0$. Thus, at round $t=1$, all correct parties multicast $\ip{\vote,u}$ and a $\forward$ message that includes all $\echo$ messages of all correct parties. In round $t=2$, no party detects equivocation from $\ell$ due to $\ell$ being correct, and all parties see a valid $\vote$ message for $u$ from all correct parties, and $H\subseteq E^{^,p}$ for all correct parties $p$. Since the adversary is $\rho$-bounded, $H$ are a strict majority of the parties, and so any correct party $p$ observes $|V_u^p\cap E^{*,p}|>\frac{|E^p|}{2}$ at round $t=2$. Thus, all correct parties decide $u$ in round $t=2$, as required.
    \end{claimproof}
This concludes the proof of \cref{thrm:BB_fast_path_unknown_participation}.
\end{proof}

\section{$\BB$ and $\BA$ with Optimal Resilience}
\label{sec:slow_path_BB_BA}

In this section, we prove the following theorems.

\begin{theorem}\label{thrm:BB_slow_path}
    For all $\rho \leq 1/2$,
    there exists a $\rho$-secure protocol (namely $\BBSP$ below) solving $\BB$ under dynamic participation with good-case latency $\Tgc=3$.
\end{theorem}

\begin{theorem}\label{thrm:BA_slow_path}
    For all $\rho \leq 1/2$,
    there exists a $\rho$-secure protocol (namely $\BASP$ below) solving $\BA$ under dynamic participation with good-case latency $\Tgc=2$.
\end{theorem}

\subsection{$\BB$ with Good-Case Latency $\Tgc=3$}
\label{ssec:BB_slow_path}

In this section, we prove \cref{thrm:BB_slow_path}.
In this protocol, we make black-box use of the $\frac{1}{2}$-secure dynamic participation $\BA$ protocol from~\cite{DACS}, which we denote by $\Pi_\BA$.

\begin{protocol}{$\BBSP$}
    \label{prot:BB_slow_path}
    \begin{enumerate}
        \item[] \textbf{Setup.} Leader $\ell$ has input $u_\ell$. Instructions for party $p$.
        \item \textbf{Round $t=0$}: $\ell$ multicasts $\ip{\echo,u_\ell}_\ell$.
        \item \textbf{Round $t=1$}: If received $\ip{\echo,u}_\ell$ for some $u$, multicast $\ip{\vote,u}_p$ along with the corresponding $\echo$ message from $\ell$. Else, multicast $\ip{\vote,\bot}_p$.
        \item \textbf{Round $t=2$}: Denote by $V$ the set of parties from which $p$ received a $\vote$ message. Multicast $\ip{\forward,V}_p$, along with the corresponding valid $\vote$ messages. 
        \item \textbf{Round $t=3$}: Denote by $G$ the set of parties from which $p$ received a $\forward$ message, and by $V$ the set of parties from which $p$ received a $\vote$ message, whether directly or via a $\forward$ message. 
        Denote by $V^*$ the set of parties $q$ for which the following holds.
        \begin{enumerate}
            \item $p$ did not detect an equivocation from $q$, i.e., two valid $\vote$ messages signed by $q$ for values $u\neq u'$.
            \item A valid $\vote$ message from $q$ appears in more than $\frac{|G|}{2}$ of the $\forward$ messages received by $p$.
        \end{enumerate}
        Let $u$ be the value $\arg\max\limits_{u\in I} |V_u\cap V^*|$. Multicast $\ip{\echotwo,V_u\cap V^*}_p$, along with the corresponding $\vote$ messages.
        
        \textbf{Early deciding condition:} If the following holds:
        \begin{enumerate}
            \item $p$ did not detect an equivocation from $\ell$, i.e., two valid $\vote$ messages signed by $\ell$ for values $u\neq u'$.
            \item $|V_u\cap V^*|>\frac{|V|}{2}$.
        \end{enumerate}
        Then decide $u$ and multicast $\ip{\decided,u}_p$.
        \item \textbf{Round $t\geq 4$}:
        \begin{enumerate}
            \item \textbf{Early deciding:} Denote by $D$ the set of parties from which $p$ received a $\decided$ message, and by $D^*$ the set of parties from which $p$ received an $\echotwo$ message. If there exists a value $u$ such that $|D_u|>\frac{|D^*|}{2}$, then decide $u$.
            \item Let $u'$ be the value for which $p$ received the largest $V_u\cap V^*$ (intersection with \emph{own} $V^*$) among all received $\echotwo$ messages. Run $\Pi_\BA$ as instructed that commences in round $t=4$ with input $u'$, and decide the output of $\Pi_\BA$, if did not decide before.
        \end{enumerate}
    \end{enumerate}
\end{protocol}

\begin{proof}[Proof of \cref{thrm:BB_slow_path}]
    Our protocol is given in $\BBSP$. We now analyze its security and efficiency. Termination follows from the termination of $\Pi_\BA$, as all parties decide and halt once $\Pi_\BA$ terminates. We next handle agreement.
    \begin{claim}\label{claim:BB_slow_path_agreement}
        $\BBSP$ satisfies agreement against any $\frac{1}{2}$-bounded adversary.
    \end{claim}
    \begin{claimproof}
        There are two cases: either there exists a correct party that decides due to the early deciding condition at round $t=3$, or not. The proof can be quickly concluded in the latter case, as then no correct party multicasts a $\decided$ message, and thus no correct party decides according to the early deciding condition in all rounds $t\geq 3$. Thus, all correct parties decide according to the output of $\Pi_\BA$, which guarantees agreement. 
        
        We are now left with the case that there exists a correct party that decides a value $u$ due to the early deciding condition at round $t=3$. First we prove that this can occur for at most a single value. Assume that there are two correct parties $p,q$ that decide according to the early deciding condition in round $t=3$ values $u\neq u'$, respectively. We thus have that $|V_u^p\cap V^{*,p}|>\frac{|V^p|}{2}$, and $|V_{u'}^q\cap V^{*,q}|>\frac{|V^q|}{2}$. In particular, both $V_u^p\cap V^{*,p}$ and $V_{u'}^q\cap V^{*,q}$ contain at least one correct party. Which implies that both $p,q$ have observed a valid $\vote$ message for both $u$ \emph{and} $u'$. This implies that both $p$ and $q$ have observed an equivocation by $\ell$, and thus would not have decided by the early deciding  condition in round $t=3$. 
        
        We are thus left with the case that there exists a unique value $u$ such that a correct party $p$ awake in round $t=3$ decides $u$ via the early deciding condition.
        First, observe that since no correct party multicasts a $\decided$ message for a value $u'\neq u$, we have that any correct party deciding via the early deciding condition in round $t\geq 4$ decides $u$ as well.
        If a correct party $p$ awake in round $t=3$ decides $u$ via the early deciding condition, this implies that $p$ did not detect an equivocation from $\ell$, and that $|V_u^p\cap V^{*,p}|>\frac{|V^p|}{2}$. Denoting by $H$ the set of correct parties awake at round $t=1$, we have that \emph{none} of them multicast a $\vote$ message for a value $u'\neq u$, as otherwise, $p$ would have detected an equivocation from $\ell$. For some correct party $q\neq p$ awake at round $t=4$, consider the set $V_{u'}^q\cap V^{*,q}$ for some value $u'\neq u$. Denote by $F$ the set of corrupt parties. As we have just observed, we have that $V_{u'}^q\cap V^{*,q}\subseteq F$. Note, however, that if some $f\in F$ satisfies $f\in V_{u'}^q\cap V^{*,q}$, then a $\vote$ message from $f$ for value $u'$ appears in a majority of the $\forward$ messages received by $q$, thus at least one of them was sent by a correct party, and thus all correct parties awake at round $t=3$, and in particular $p$, received a valid $\vote$ message for $u'$. This contradicts the assumption that $p$ decided the value $u$ via the early deciding condition at round $t=3$. We thus have that $V_{u'}^q\cap V^{*,q}=\emptyset$ for all correct parties awake at round $t\geq 4$. This in particular implies that all correct $q$ parties awake at round $t=4$ observe $u$ to be the value maximizing $V^q_u\cap V^{*,q}$ among all $\echotwo$ messages, and commence $\Pi_\BA$ with input $u$. This triggers the validity property of $\Pi_\BA$ which implies that all correct parties output $u$ from $\Pi_\BA$, and thus all correct parties decide $u$, as required.
    \end{claimproof}

    Next, we handle validity and good-case latency.
    \begin{claim}\label{claim:BB_slow_path_validity}
        $\BBSP$ satisfies validity against any $\frac{1}{2}$-bounded adversary. Furthermore, $\BBSP$ has good-case latency 3.
    \end{claim}
    \begin{claimproof}
        Assume $\ell$ is correct with input $u$. Then by behavior of $\BBSP$, it multicasts $\ip{\echo,u}_\ell$ to all parties in round $t=0$. Afterwards, again by protocol behavior, all correct parties awake in round $t=1$, denoted by $H$, multicast a valid $\ip{\vote,u}_p$ message. In round $t=2$, all correct parties multicast $\forward$ messages containing all vote messages of all correct parties awake in round $t=1$. In round $t=3$, note first that no correct party $p$ detects an equivocation from $\ell$, due to $\ell$ being correct. Second of all, note that $H\subseteq V_u^p\cap V^{*,p}$, and $|V|\leq |H|+|F|$. Thus, we have that $|V_u^p\cap V^{*,p}|>\frac{|V|}{2}$ holds for any correct party $p$ awake in round $t=3$. Thus, all correct parties awake in round $t=3$ decide $u$ and multicast a $\decided$ message for $u$. An identical argument then gives that all correct parties decide $u$ in the first round $t\geq 4$ in which they are awake, due to the early deciding condition.
    \end{claimproof}

    This concludes the proof of \cref{thrm:BB_slow_path}.
\end{proof}

\subsection{$\BA$ with Good-Case Latency $\Tgc=2$}
\label{ssec:BA_slow_path}

In this section, we prove \cref{thrm:BA_slow_path}. 
In this protocol, we make black-box use of the $\BA$ protocol from~\cite{DACS}, which we denote by $\Pi_\BA$, and which is $\frac{1}{2}$-secure under dynamic participation. Throughout the protocol, whenever we consider a set of messages $S$, the notation $S^p$ refers to the local view, or the purported local view, of $p$ of the set $S$.

\begin{protocol}{\BASP}
    \label{prot:BA_slow_path}
    \begin{enumerate}
        \item[] \textbf{Setup.} Each party $p$ has input $u_p$. Instructions for party $p$.
        \item \textbf{Round $t=0$}: Multicast $\ip{\echo,u_p}_p$.
        \item \textbf{Round $t=1$}: Denote by $E$ the set of parties from which $p$ received an $\echo$ message. Multicast $\ip{\forward,E}_p$, along with the corresponding $\echo$ messages.
        \item \textbf{Round $t=2$}: Denote by $G$ the set of parties from which $p$ received a $\forward$ message. Denote by $E$ the set of parties from which $p$ received an $\echo$ message, either directly or via a $\forward$ message. For a value $u$, denote by $E_u$ the set of parties from which $p$ received an $\echo$ message for $u$. Denote by $E^*$ the set of parties $q$ for which the following holds.
        \begin{enumerate}
            \item $p$ detected no equivocation from $q$, i.e., two $\echo$ messages signed by $q$ for values $u'\neq u$.
            \item An $\echo$ message from $q$ appears in more than $\frac{|G|}{2}$ of the $\forward$ messages received by $p$
        \end{enumerate}
        For a value $u$, denote by $E^{*}_u$ the set $E_u\cap E^*$.
        
        \textbf{Early deciding condition:} If there exists a value $u$ such that $|E^*_u|>\frac{|E|}{2}$, then decide $u$. Multicast $\ip{\decided,u, E,E_u^*,\set{\forward_q\mid q\in G}}$, where $\forward_q$ indicates the $\forward$ message $p$ received from $q$.
        Else, multicast $\ip{\echotwo,E,E^*,\set{\forward_q\mid q\in G}}$.
        \item \textbf{Round $t\geq3$}.
        \begin{itemize}
            \item Denote by $E$ the set of parties from which $p$ received an $\echo$ message, and by $E_u$ the set of parties from which $p$ received an $\echo$ message for $u$ .
            \item Denote by $E^*$ the set defined in an identical manner as in round $t=2$ above. 
            \item Denote by $G$ the set of parties from which $p$ received a $\forward$ message. 
            \item Denote by $T$ the set of parties from which $p$ received an $\echotwo$ message. 
            \item Denote by $G^*$ the set of parties $q$ for which a $\forward$ message from $q$ appeared in more than $\frac{|T|}{2}$ of the $\echotwo$ messages. 
        \end{itemize}  
        \begin{enumerate}
            \item \textbf{Early deciding:} For a value $u$, denote by $D_u$ the set of parties from which $p$ received a $\decided$ message for value $u$. If there exists a value $u$ such that $|D_u|>\frac{|T|}{2}$, then decide $u$.
            \item For a party $p$, we say a $\decided$ message from a party $q$ of the form $\ip{\decided, u,E^q,E_u^{*,q},\set{\forward_z\mid z\in G^q}}$ is \emph{valid} if the following holds.
            \begin{itemize}
                \item $E^q$ contains $E^{*,p}$.
                \item $G^q$ contains $G^{*,p}$.
                \item Each $z\in E_u^{*,q}$ appears in a majority of $G^q$, and there are no equivocations from $z$ in $\set{\forward_y\mid y\in G^q}$.
                \item $|E_u^{*,q}|>\frac{|E^q|}{2}$.
            \end{itemize}

           For a valid $\decided$ message for a value $u$ from a party $q$, we refer to $|E^q|$ as its \emph{size}. Let $u'$ be the value for which $p$ received the maximum size $\decided$ message. Run $\Pi_\BA$ as instructed that commences in round $t=3$ with input $u'$, and decide the output $\Pi_\BA$, if did not decide beforehand.
        \end{enumerate}

    \end{enumerate}
\end{protocol}

\begin{proof}[Proof of \cref{thrm:BA_slow_path}]
    Our protocol is given in $\BASP$. We now analyze its security and efficiency. Termination follows from the termination of $\Pi_\BA$, as all parties decide and halt once $\Pi_\BA$ terminates. We next handle agreement.

    \begin{claim}\label{claim:BA_slow_path_agreement}
        $\BASP$ satisfies agreement against any $\frac{1}{2}$-bounded adversary.
    \end{claim}
    \begin{claimproof}
        There are two cases: either there exists a correct party that decides due to the early deciding condition at round $t=2$, or not. The proof can be quickly concluded in the latter case, 
        as all correct parties decide according to the output of $\Pi_\BA$, which is $\frac{1}{2}$-secure and guarantees agreement.
        
        We are now left with the case that there exists a correct party that decides a value $u$ due to the early deciding condition at round $t=2$. First, we prove that this can occur for at most a single value. Assume that two correct parties $p,q$ decide $u,u'$ according to the early deciding condition in round $t=2$, respectively. In particular, that implies that $|E_u^{*,p}|>\frac{|E^p|}{2}$, and $|E_{u'}^{*,q}|>\frac{|E^{q}|}{2}$. Note that by the behavior of the protocol we have that  $E^{*,p},E^{*,q}\subseteq E^{p}$, and similarly $E^{*,p},E^{*,q}\subseteq E^{q}$, since every echo in $E^{*,p},E^{*,q}$ appeared in the $\forward$ message of at least one correct party. 
        Denote by $F_u^p$ the set of corrupt parties included in $E_u^{*,p}$, and by $F_{u'}^q$ the set of corrupt parties included in $E_{u'}^{*,q}$. Note that $E_u^{*,p}\cap E_{u'}^{*,q}=\emptyset$, as otherwise $p,q$ would have detected an equivocation from the party in the intersection, and would not have included it in their $E^*$ set.

        Denote by $H^0$ the set of correct parties awake at round $t=0$, and of those, by $H^0_u,H^0_{u'}$ those with input $u,u'$, respectively.
        One has that
        \[        |H^0_u|+|F^p|=|E_u^{*,p}|>\frac{|E^{p}|}{2}\geq \frac{|H^0|+|F^p|+|F^q|}{2}
        \]
        Similarly, one has that
        \[     |H^0_{u'}|+|F^q|=|E_{u'}^{*,q}|>\frac{|E^{q}|}{2}\geq \frac{|H^0|+|F^p|+|F^q|}{2}
        \]
        Adding these inequalities, one gets that $|H^0_u|+|H^0_{u'}|>|H^0|$, which is a contradiction.
        
        We are then left with the case that there exists a unique value $u$ such that a correct party $p$, decides $u$ at round $t=2$, with reported participation $E^{p}$. First note that this implies that any correct party that decides due to the early deciding condition in round $t\geq 3$ also decides $u$, as no correct party multicasts a $\decided$ message for any value $u'\neq u$.
        Now, assume that some correct party $q$, awake in round $t=3$, received a $\decided$ message from a party $z\in F$ for some value $u'$, of the form $\ip{\decided,u',E^z,E_{u'}^{*,z},\set{\forward_s\mid s\in G^z}}$, with reported participation $E^{z}$, such that $|E^{z}|\geq |E^{p}|$. We now prove this cannot happen. Assume otherwise (i.e. $|E^{z}|\geq |E^{p}|$), and denote by $E_{u'}^{*,z}$ the claimed such set of $z$.
        Denote by $F^p_u$ the set of corrupt parties in $E_u^{*,p}$. We have that $|E_u^{*,p}|=|H^0_u|+|F^p_u|$. 
        We next note that $E^{p}$ contains the 
         set 
         $E_{u'}^{*,z}$. This holds due to the following: A party $s\in E_{u'}^{*,z}$ must appear in a strict majority of $\forward$ messages reported by $z$, i.e., the set $G^z$. Since we assume that $z$ sends a \emph{valid} $\decided$ message, we have that $G^{*,q}\subseteq G^z$. Denoting by $H^1$ the set of correct parties awake at round $t=1$, we have that $H^1\subseteq G^{*,q} \subseteq G^z$. Thus, if party $s$ appears in a strict majority of the $\forward$ messages of the claimed set $G^z$, then at least one of these $\forward$ messages was sent by a correct party at round $t=1$. Thus, $p$, which was awake at time $t=2$, observes an $\echo$ message from $s$ for $u'$, and thus $s\in E^p$. This gives $E_{u'}^{*,z}\subseteq E^p$. Denoting by $F_{u'}^z$ the corrupt parties in $E_{u'}^{*,z}$, we in particular have $F_{u'}^z\subseteq E^p$.
         
         In fact, note that this line of arguing actually implies a stronger property, and that is that $E_{u'}^{*,z}\cap E^{*,p}_u=\emptyset$, as $p$ seeing an echo for $u'$ from $s$ implies that $s\not\in E_u^{*,p}$.
        From $F_{u'}^z\subseteq E^p$, and $E_{u'}^{*,z}\cap E^{*,p}_u=\emptyset$, we can deduce that: $|E^{*,p}_u|=|H^0_u|+|F^p_u|>\frac{|E^p|}{2}\geq \frac{|F_{u'}^{z}|+|F^p_u|+|H^0|}{2}$. 
        We now have all we need to arrive at a contradiction. Let us summarize our observations.
        \begin{itemize}
            \item $E_{u'}^{*,z}\cap E^{*,p}_u=\emptyset$ which in particular implies that $F_{u'}^{z}\cap F_u^{p}=\emptyset$.
            \item $|E^{*,p}_u|=|H^0_u|+|F^p_u|>\frac{|E^p|}{2}\geq \frac{|F_{u'}^{z}|+|F^p_u|+|H^0|}{2}$.
            \item $|E_{u'}^{*,z}|\leq |H_{u'}|+|F_{u'}^z|$. This bound trivially holds.
            \item $|E^{z}|\geq |E^{p}|$, our assumption.
            \item $|E_{u'}^{*,z}|>\frac{|E^z|}{2}$, from the assumption that $z$'s $\decided$ message is valid.
        \end{itemize}
        From these observation we deduce that:
        \[|H^0_u|+|H^0_{u'}|+|F^{p}_u|+|F^z_{u'}|\geq |E_{u'}^{*,z}|+|E_u^{*,p}|>|E^{p}|\geq |H^0|+|F^{p}_u|+|F^z_{u'}|
        \]
        This finally implies that $|H^0_u|+|H^0_{u'}|>|H^0|$, which is a contradiction. We thus have that all correct parties awake in round $t=3$ view the $\decided$ message from $p$ as the valid  $\decided$ message with the maximum size ($|E^p|$), and so they all commence $\Pi_\BA$ with input $u$. This triggers the validity property of $\Pi_\BA$, and thus all correct parties output $u$ from $\Pi_\BA$ and decide $u$, and agreement holds.
    \end{claimproof}

Next, we handle validity and good-case latency.
    \begin{claim}\label{claim:BA_slow_path_UP_validity}
        $\BASP$ satisfies validity against any $\frac{1}{2}$-bounded adversary. Furthermore, $\BASP$ has good-case latency $2$.
    \end{claim}
    \begin{claimproof}
        Assume that all correct parties awake in round $t=0$ have the same input $u$. Denoting by $H^0$ the set of correct parties awake in round $t=0$ and by $F$ the set of corrupt parties, we thus have that $E_u^{*,p}$ contains $H^0$ for all correct parties $p$, as they all send an $\echo$ message for $u$ in round $t=0$, and none of them equivocate. It clearly holds, for every party $p$, that $|E^{p}|\leq |H|+|F|$, and since $|H|>|F|$ we thus have that $|E^{*,p}_u|>\frac{|E^{p}|}{2}$ for all correct parties $p$ in round $t=2$. Thus, all correct parties awake in round $t=2$ (denoted by $H^2$) decide $u$ and multicast a $\decided$ message for $u$. Any correct party $q$, in the first round $t\geq 3$ in which it is awake, observes $\decided$ messages for $u$ from all parties in $H^2$. This constitutes a majority of the received $\echotwo$ messages, and so any correct party $q$ also decides $u$ in the first round $t\geq 3$ in which it is awake.
    \end{claimproof}
    This concludes the proof of \cref{thrm:BA_slow_path}.
\end{proof}

\begin{credits}
\subsubsection*{Acknowledgment}
We thank
David Tse
for fruitful discussions.
The work of YE was conducted in part while at a16z Crypto Research.
\end{credits}

\bibliographystyle{plainurl}
\bibliography{references}

\appendix
\section{Folklore BA for Static Participation with $\rho<1/3$ and $\Tgc=1$}
\label{sec:one_round_sync_BA}

In this section we show, for completeness, the following folklore results.
\begin{lemma}
    \label{lemma:one_round_sync_BA}
    For all $\rho \leq 1/3$,
    there exists a $\rho$-secure $\BA$ protocol (namely $\Sync_{\BA,1}$ below) with good-case latency $\Tgc=1$ under static participation in synchrony.
\end{lemma}

\begin{lemma}
    \label{lemma:one_round_sync_BA_lower_bound}
    For all $\rho>\frac{1}{3}$,
    there is no $\rho$-secure $\BA$ protocol with good-case latency $\leq 1$ under static participation in synchrony.
\end{lemma}

\begin{proof}
    Denote by $\Pi_\BA$ an arbitrary choice of $\BA$ protocol which is $\frac{1}{3}$-secure under static participation in synchrony. We treat it as a black-box. Our protocol is given below:
    \begin{protocol}{$\Sync_{\BA,1}$}
        \begin{enumerate}
            \item[] \textbf{Setup.} Each party $p$ has input $u_p$. Instructions for party $p$. Denote by $n$ the number of parties and by $f<\frac{n}{3}$ the bound on the number of corrupted parties. Initialize $o_p \leftarrow u_p$.
            \item \textbf{Round} $t=0$: Multicast $\ip{\echo,u_p}_p$.
            \item \textbf{Round} $t=1$: \textbf{Early deciding condition.} If there exists a value $u$ such that $p$ received $n-f$ $\echo$ messages for $u$, then decide $u$. Denote $n-f$ $\echo$ messages for a value $u$ by $\cC(u)$. Multicast $\ip{\decided,\cC(u)}$.
            \item \textbf{Round $t\geq 2$:} If received $\cC(u)$ for some value $u$, set $o_p \leftarrow u$. Run $\Pi_\BA$ with input $o_p$ that commences at round $t=2$, and decide accordingly, if have not decided before.
        \end{enumerate}
    \end{protocol}

    Termination is clear from the termination of $\Pi_\BA$. For agreement, if no correct party decides via the early deciding condition in round $t=1$, then agreement holds by the agreement property of $\Pi_\BA$. Else, let $p$ be a correct party that decides a value $u$ in round $t=1$, thus $p$ observed $n-f$ $\echo$ messages for $u$, since $f<\frac{n}{3}$, a quorum intersection argument implies that no $n-f$ $\echo$ messages exist for any other value, and so all correct parties set their local $o$ variable to $u$ in round $t=2$. This triggers the validity property of $\Pi_\BA$, and so all correct parties output $u$, as required. For validity and good-case latency, consider the case where all correct parties have the same input value $u$. Then by the behavior of the protocol, all correct parties see in round $t=1$ $n-f$ $\echo$ messages for $u$, and decide $u$, as required. This concludes the proof.
\end{proof}
    
Next, we prove \cref{lemma:one_round_sync_BA_lower_bound}.
\begin{proof}
    Let $\rho>\frac{1}{3}$, and let $n$ be a natural number, and consider three disjoint sets of parties $A,B,C$, each of size $\frac{n}{3}$. Assume towards a contradiction that there exists a $\rho$-secure protocol solving $\BA$ under synchrony with static participation and good-case latency $1$. We consider three executions $G_1,G_2,G_3$, as follows.
    \begin{enumerate}
        \item $G_1$: $A,B$ are correct with input $1$. $C$ are corrupt, and behave as if they are correct with input $0$. Note that the adversary is $\rho$-bounded.
        \item $G_2$: $A,C$ are correct, $B$ are corrupt. The adversary is $\rho$-bounded. $A$ has input $1$, $C$ has input $0$. $B$ behave in round $t=0$ to $A$ as if they are correct with input $1$, and to $C$ as if they are correct with input $0$, $B$ then halts.
        \item $G_3$: Dual to $G_1$. $B,C$ are correct with input $0$. $A$ are corrupt, and behave as if they are correct with input $1$. Note that the adversary is $\rho$-bounded.
    \end{enumerate}

    Note that in $G_1$, all correct parties have the same input $1$, and so the parties in $A$ decide $1$ after one round. Note that up to the end of the first round, $G_1,G_2$ are indistinguishable from the perspective of $A$, and so also in $G_2$, $A$ decides $1$ after $1$ round. A dual argument gives that $C$ outputs $0$ after one round both in $G_3$ and $G_2$. Lastly note that in $G_2$, the adversary is $\rho$-bounded, and so by assumption that $\Pi$ is $\rho$-secure, agreement must hold and $A,C$ output the same value in $G_2$, leading to a contradiction.
\end{proof}

\end{document}